\documentclass[11pt,a4paper,headinclude,footinclude,chapterprefix=on]{scrreprt}

\usepackage{tabularx}
\usepackage{multicol}
\usepackage{graphicx}
\usepackage{acronym}
\usepackage{url}

\usepackage{tikz}
\usepackage{lipsum}

\usepackage[headsepline,plainheadsepline,footsepline,plainfootsepline]{scrpage2}

\usepackage[scaled=.92]{helvet}
\usepackage{courier}

\usepackage[multiuser,nomargin,marginclue,footnote]{fixme}
\fxusetheme{color}
\fxsetup{targetlayout=colorcb}
\FXRegisterAuthor{awo}{anawo}{AWO}
\fxsetup{status=draft}

\usepackage{geometry}
\usepackage[pscoord]{eso-pic}
\newcommand{\placetextbox}[3]{
  \setbox0=\hbox{#3}
  \AddToShipoutPictureFG*{
    \put(\LenToUnit{#1\paperwidth},\LenToUnit{#2\paperheight}){\vtop{{\null}\makebox[0pt][l]{#3}}}%
  }%
}%

\usepackage{calc}
\usepackage{upgreek}
\usepackage{amsmath,amssymb,amsfonts,amsbsy,amsthm}
\usepackage{mathtools}
\usepackage{bbm}

\clearscrheadfoot

\ihead[\Large {\scshape TU Berlin }]{\Large {\scshape TU Berlin }}

\ifoot[{\tiny \begin{minipage}{5cm}Copyright at Technische Universit\"at Berlin.\newline All rights reserved.\end{minipage}}]{{\tiny \begin{minipage}{5cm}Copyright at Technische Universit\"at Berlin.\newline All rights reserved.\end{minipage}}}
\ofoot[Page \pagemark]{Page \pagemark}

\addtokomafont{pagefoot}{\normalfont}

\usepackage[noend]{algpseudocode}
\algnewcommand\algorithmicassert{\textbf{assert}}
\algnewcommand\Assert[1]{\State{}\algorithmicassert\;\;#1}
\algnewcommand\algorithmicinput{\textbf{Input:}}
\algnewcommand\Input{\item[\algorithmicinput]}
\algnewcommand\algorithmicoutput{\textbf{Output:}}
\algnewcommand\Output{\item[\algorithmicoutput]}

\makeatletter
\algrenewcommand\algorithmiccomment[2][\normalsize]{{#1\hfill\(\triangleright\) \small{\textit{#2}}}}
\makeatother

\newcommand{\trnumber}{TKN-16-001}
\newcommand{\trdate}{March 2016}
\newcommand{\trauthor}{Konstantin Miller, Abdel-Karim Al-Tamimi,\\and Adam Wolisz}
\newcommand{\tremail}{\{miller,wolisz\}@tkn.tu-berlin.de, altamimi@yu.edu.jo}
\newcommand{\trtitle}{QoE-Based Low-Delay Live Streaming Using Throughput Predictions}

\begin{document}


{
\sffamily

\thispagestyle{empty}

\setlength{\tabcolsep}{0pt}
\noindent
\begin{tabularx}{\columnwidth}{cXc}
  \includegraphics[height=1cm]{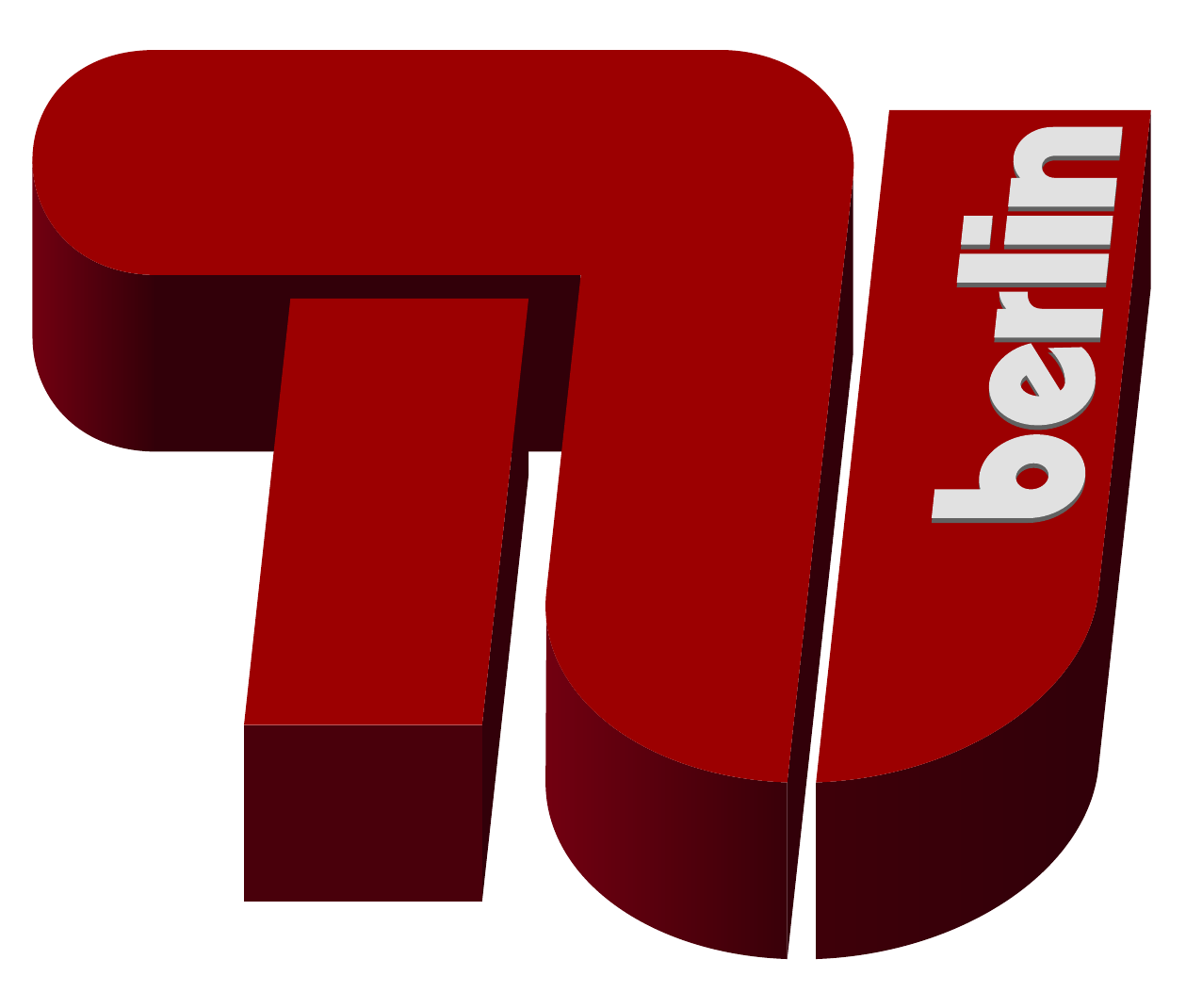}
  & &
  \includegraphics[height=1cm]{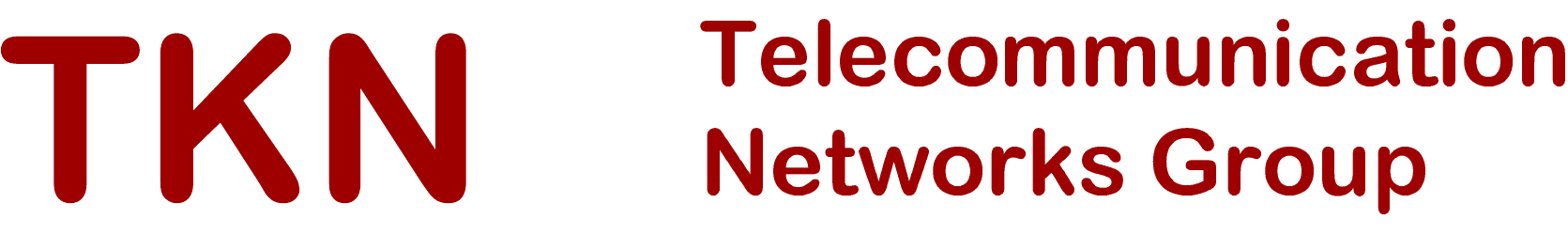}
  \\
\end{tabularx}
\setlength{\tabcolsep}{6pt}

\vspace{1.0cm}

\begin{center}
{\huge
\noindent
Technische Universit\"at Berlin

\vspace{0.5cm}

\noindent
Telecommunication Networks Group

\begin{center}
\rule{15.5cm}{0.4pt}
\end{center}
}
\end{center}

\begin{minipage}[][11.0cm][c]{14.5cm}
{\Huge

\begin{center}
\trtitle*
\end{center}
\placetextbox{0.15}{0.05}{*~This technical report updates TR TKN-15-001.}%

\begin{center}
{\LARGE \trauthor} \\
{\Large \tremail}
\end{center}

\begin{center}
Berlin, \trdate
\end{center}

\vspace{0.5cm}

}

\begin{center}
\setlength{\fboxrule}{2pt}\setlength{\fboxsep}{2mm}
\fbox{TKN Technical Report \trnumber}
\end{center}

\end{minipage}

\setlength{\fboxrule}{0.4pt}
\setlength{\fboxsep}{0.4pt}

\begin{center}

  \rule{15.5cm}{0.4pt}

  \vspace{0.5cm}

  {\huge {TKN Technical Reports Series}}

  \vspace{0.5cm}

  {\huge Editor: Prof. Dr.-Ing. Adam Wolisz}

  \vspace{0.5cm}

 \end{center}

}

\begin{abstract}
\section*{\abstractname}

Recently, HTTP-based adaptive streaming has become the \textit{de~facto} standard for video streaming over the Internet. It allows clients to dynamically adapt media characteristics to varying network conditions in order to ensure a high quality of experience, that is, minimize playback interruptions, while maximizing video quality at a reasonable level of quality changes. In the case of live streaming, this task becomes particularly challenging due to the latency constraints. The challenge further increases if a client uses a wireless access network, where the throughput is subject to considerable fluctuations. Consequently, live streams often exhibit latencies of up to 20 to 30 seconds. In the present work, we introduce an adaptation algorithm for HTTP-based live streaming called LOLYPOP (\textbf{Lo}w-\textbf{L}atenc\textbf{y} \textbf{P}redicti\textbf{o}n-Based Ada\textbf{p}tation) that is designed to operate with a transport latency of few seconds. To reach this goal, LOLYPOP leverages TCP throughput predictions on multiple time scales, from 1 to 10 seconds, along with an estimate of the relative prediction error distribution. In addition to satisfying the latency constraint, the algorithm heuristically maximizes the quality of experience by maximizing the average video quality as a function of the number of skipped segments and quality transitions. In order to select an efficient prediction method, we studied the performance of several time series prediction methods in IEEE 802.11 wireless access networks. We evaluated LOLYPOP under a large set of experimental conditions, limiting the transport latency to 3 seconds, against a state-of-the-art adaptation algorithm from the literature, called FESTIVE\@. We observed that the average video quality is by up to a factor of 3 higher than with FESTIVE\@. We also observed that LOLYPOP is able to reach a broader region in the quality of experience space, and thus it is better adjustable to the user profile or service provider requirements.

\end{abstract}

\tableofcontents

\chapter{Introduction}

Over the last few years, we have been observing a dramatic change in video consumption patterns. The era of passive consumption of non-interactive ``linear'' content on a single device, the TV set, appears to be coming to an end, and a new mindset is being established: watch what I want, when I want, and where I want~\cite{comScore2014a,Conviva2015}.
This transformation is accompanied by a significant increase in wireless and mobile network usage. In 2013, wired devices still accounted for the majority of Internet traffic at 56\%.
However, wireless and mobile device traffic is predicted to exceed traffic from wired devices by 2018, accounting for 61\% of the total Internet traffic. The largest part of it will be video content~\cite{CiscoVNI2014}.

Although the majority of streamed content is \ac{VoD}, the amount of live streaming is growing rapidly~\cite{Gigaom2014}. While current live streaming services can exhibit a latency of several tens of seconds, \emph{low-delay} streaming refers to live streaming with a particularly low upper bound on the latency: a few seconds or less. Such a requirement is desirable for scenarios such as transmissions of sports events. Moreover, a low latency is absolutely necessary in the case of video conferencing and online gaming, where active participants have latency requirements on the order of hundreds of milliseconds~\cite{ITU-T_F.746.1}, while permanently or temporarily passive participants may be served with a delay of few seconds.

Streaming video over the Internet has always been a challenging task because the Internet was not designed to support applications that require guaranteed end-to-end \ac{QoS}~\cite{ITU-T_E.800}. Even though considerable effort has been put into developing networking architectures addressing this shortcoming~\cite{Aurrecoechea1998,Carapinha2010}, none of the approaches have achieved a significant pervasiveness. As a result, in 2013 around 26.9\% of streaming sessions on the Internet experienced playback interruption, 43.3\% were impacted by low resolution, and 4.8\% failed to even start altogether~\cite{Conviva2014}. Especially on wireless links, users are exposed to interference, cross-traffic, and fading effects, leading to continuously fluctuating \ac{QoS} characteristics.

As a consequence, we lately have been observing a period of high interest in adaptive streaming technologies that are able to dynamically adjust the characteristics of the streamed media to varying network conditions, leading to a smoother viewing experience with less playback interruptions and a more efficient utilization of the available network resources. In particular, one technology has become the \textit{de~facto} standard for Internet streaming: \ac{HAS}~\cite{Stockhammer2011a}. One of the enablers of its success was the open standard MPEG-DASH (Dynamic Adaptive Streaming over HTTP)~\cite{DASH2012,Sodagar2011}. The advantage of \ac{HAS} comes from its usage of \ac{HTTP} leveraging an ubiquitous and highly optimized delivery infrastructure, including \acp{CDN}, caches, and proxies, reducing the operating costs due to the lack of necessity to maintain specialized video servers and pay for their licenses. In addition, \ac{HTTP} is typically allowed to traverse middleboxes, such as \ac{NAT} devices and firewalls. Finally, \ac{HAS} has good scalability properties due to the stateless nature of \ac{HTTP}, and since the control logic resides at the client.

\ac{HAS}, however, was primarily developed to replace the progressive download of \ac{VoD} content and therefore its usage for low-delay streaming has received little attention in the research community. Typical buffer sizes used in studies for evaluation and deployment of \ac{HAS}-based clients are on the order of tens of seconds. The capability of the \ac{HAS} approach to efficiently stream low-delay content, especially in wireless networks, is still an open question.

One of the main goals of a streaming client's adaptation logic is to maximize \ac{QoE}. The notion of \ac{QoE} has been introduced in an effort to make the various phenomena affecting human perception of multimedia content accessible to an objective evaluation process~\cite{ITUT_P10G100_Am2}. Among the \ac{QoE} influencing factors that are controlled by the adaptation logic are the number of playback interruptions, the number of quality transitions, and the video quality. It is worth noting that these three factors cannot be considered separately. For example, always selecting the lowest video quality minimizes playback interruptions and quality transitions and allows for the lowest playback latency. On the other hand, always selecting the highest video quality typically results in an unacceptably high number of playback interruptions.

In this study, we demonstrate that efficient \ac{HAS}-based low-delay live streaming is possible by leveraging short-term \ac{TCP} throughput predictions over multiple time scales, from 1 to 10 seconds, along with estimations of the relative prediction error distribution. We design a novel prediction-based algorithm called \ac{LOLYPOP} that supports quality-based adaptation with a transport latency on the order of a few seconds.
The approach introduced in \ac{LOLYPOP} jointly considers three \ac{QoE} components: the number of playback interruptions, the number of quality transitions, and the average video quality. Its goal is to maximize the average video quality as a function of the operating point defined by the other two components. The operating point is controlled by two input parameters: an upper bound on the number of quality transitions and a parameter controlling the number of playback interruptions. Thus, \ac{LOLYPOP} provides configurable \ac{QoE} that can be adjusted to the nature of the video, the user context and preferences, or the service provider's business model.


At the core of \ac{LOLYPOP} is an estimation of download success probabilities for individual segments. To obtain these estimations, \ac{LOLYPOP} leverages predictions of throughput distributions, computed from a time series prediction and an error estimation. We evaluate several time series prediction methods using \ac{TCP} throughput traces collected in IEEE 802.11 \acp{WLAN}, including public hotspots (indoor and outdoor), campus hotspots, and access points in residential environments. We observe, somewhat surprisingly, that taking the average over the previous $T$ seconds as a prediction for the next $T$ seconds provides the best prediction accuracy among the considered methods for all considered time scales. That is, taking into account the trend does not help to reduce the prediction error.

We implement a prototype of the algorithm and evaluate it against FESTIVE~\cite{Jiang2014}, a well-known adaptation algorithm from the literature. We limit the transport latency to 3 seconds using a segment duration of 2 seconds. We observe that LOLYPOP is able to reach a broad range of operating points and thus can be flexibly adapted to the user profile or service provider requirements. Furthermore, we observe that at the individual operating points, \ac{LOLYPOP} provides an average video quality which is by up to a factor of 3 higher than the quality achieved by the baseline approach.

The rest of the paper is structured as follows. After reviewing the related work in Chapter~\ref{sec:related_work}, we introduce the proposed system's model and used notation in Chapter~\ref{sec:system_model}. Afterwards, we introduce the design of the proposed adaptation algorithm in Chapter~\ref{sec:adaptation}. After describing our collected set of throughput traces in Chapter~\ref{sec:traces}, we use them to compare the performance of several prediction methods in Chapter~\ref{sec:prediction}, where we also present our approach to computing download success probabilities. We evaluate the performance of \ac{LOLYPOP} in Chapter~\ref{sec:evaluation}, and present our conclusions in Chapter~\ref{sec:conclusion}.

\chapter{Related Work}
\label{sec:related_work}

In the last few years, adaptive streaming has been a very active research area. Even though the idea is not new~\cite{Chen1995,Kanakia1995}, it has recently moved into focus as the number of streaming video services has increased as well as the number of content providers using adaptive streaming on a large scale. \ac{HTTP} was considered as a candidate application-layer streaming protocol as early as 2002~\cite{Carmel2002} but, it was not until recently that it became the technology of choice for delivering video over the Internet~\cite{Swaminathan2013,Li2013a}. The adoption of the open standard MPEG-DASH (Dynamic Adaptive Streaming over HTTP)~\cite{DASH2012,Sodagar2011} in 2011 has significantly contributed to the popularity of \ac{HAS}.

There is a large number of recent studies focusing on adaptation algorithms for \ac{HAS} that do not address the low-delay requirement. They typically consider playback buffer sizes of 10 to 30 seconds or more. Various approaches have been proposed that are based on control theory~\cite{Zhou2012,Zhu2013,Zhou2014,Yin2014,Miller2015}, Markov decision processes~\cite{Jarnikov2011a,Bokani2013}, machine learning~\cite{Claeys2014}, data mining~\cite{Balachandran2013}, dynamic programming~\cite{Li2014}, data-driven techniques~\cite{Liu2012a,Riiser2012,Hao2014}, and other heuristics selecting video quality based on the client's playback buffer level and average throughput~\cite{Liu2011,Mok2012,MillerK2012,Li2013,Jiang2014}. Several studies use cross-layer information to improve client's performance~\cite{Abdallah2015}, or jointly consider the problem of video quality selection and resource allocation on the lower layers of the \ac{OSI} model~\cite{Miller2015,Le2015a,Bethanabhotla2015}. Further studies propose to coordinate the quality selection process among the clients sharing a bottleneck link to allow for a more fair and efficient resource usage~\cite{Essaili2013}. Finally, many popular streaming platforms such as YouTube\footnote{http://www.youtube.com} or Netflix\footnote{http://www.netflix.com}, are using \ac{HAS} for their \ac{VoD} services, often deploying proprietary adaptation algorithms. Unfortunately, due to the lack of a standard evaluation methodology and performance metrics for \ac{HAS} systems, the results of individual studies are hard to compare with each other.

In contrast, the number of studies that specifically target live or low-delay \ac{HAS} is significantly smaller. A potential reason is the significantly smaller market size for \ac{OTT} live streaming services~\cite{Gigaom2014} and a popular opinion that streaming over HTTP/TCP is less suitable for applications requiring a low delay. In~\cite{Lohmar2011}, the authors compare the delay and communication overhead in \ac{HTTP}-based live streaming with the one in \ac{RTP}-based systems. They observe that the transport delay is by one segment duration larger with \ac{HTTP} and that the communication overhead becomes substantial for sub-second segment durations (approx. 31 kbps for one second segments). In~\cite{Thang2014}, several adaptation algorithms are evaluated using buffer sizes from 6 to 20 seconds. The authors observe that the used set of representations can significantly affect the performance of individual methods, resulting in by up to a factor of 2.8 higher average video quality. They also observe that none of the selected five methods performs best in all cases but that the ranking depends on the throughput variability. However, the study was performed using only one throughput trace. In~\cite{Wei2014}, the authors demonstrate that the server push feature introduced in \ac{HTTP}/2 can be used to support low-delay streaming by allowing a reduction in the segment duration, and thus the lower bound on the achievable delay, without suffering from the super-linear increase in the number of requests observed with \ac{HTTP}/1.1. However, the study does not analyze the protocol overhead caused by response headers and the reduction in video compression rate due to the decreased \ac{GOP} size. In~\cite{Le2015}, an adaptation algorithm for live streaming is proposed that compares potential adaptation trajectories for the next few segments and heuristically maximizes the video quality represented by the \ac{JND} metric. The algorithm is evaluated using a maximum buffer size of 10 seconds and compared against two baseline approaches. The evaluation shows that the algorithm exhibits lower average video quality than the maximum of the two baseline methods. However, it is able to reduce the average step size of a quality transition by 32\% w.r.t.\ the minimum of the two baseline methods. The performed evaluation used only one throughput trace.

Several studies assume perfect information about future throughput to compute optimal adaptation trajectories that can be used to benchmark existing algorithms and evaluate the potential for performance increase~\cite{Miller2013a,Zou2015}. The evaluation in~\cite{Miller2013a} reveals that the tested streaming clients can achieve a performance that is close to optimum w.r.t.\ the average media bit rate and number of playback interruptions, however, with an average number of quality transitions that is at least several times as high as the optimum. The evaluated buffer sizes vary between 10 and 40 seconds.

There exist other studies that explicitly use throughput predictions in the context of video quality adaptation. In \cite{Mok2012}, the authors consider path probing techniques to obtain throughput estimations that, however, typically requires support from the network infrastructure, server instrumentation, and/or modifying lower protocol layers. The proposed adaptation algorithm is not evaluated. Similar in spirit to our work is the study in~\cite{Liu2014a}, which uses predictions to match a target number of skipped segments and a minimum delay between quality transitions to control the transitions frequency. The algorithm is evaluated in a mobile network and compared against two baseline approaches. The evaluation reveals that the baseline approaches, using default configurations, require up to an order of magnitude more quality transitions in order to achieve a similar media bit rate and a similar number of skipped segments as the proposed algorithm. Unfortunately, only one network environment is used for the evaluation, and the characteristics of the observed throughput dynamics are not disclosed in the study. Also, the maximum buffer size used in the evaluation is not specified. In \cite{Yin2014}, the authors study the effect of prediction errors on the performance of two adaptation approaches, rate based and \ac{MPC}, compared against a buffer based approach using synthetic throughput traces. The authors conclude that when the prediction error exceeds 25\%, prediction-based approaches might exhibit worse performance than the buffer based algorithm. The used \ac{MPC} algorithm is not described in the publication. The study in~\cite{Tian2012} proposes a prediction-based adaptation algorithm, where the media bit rate is selected to equal the predicted throughput times a dynamically varying adjustment factor. The proposed approach is, however, not designed to support \ac{QoE}-based performance targets and has a set of configuration parameters that does not allow for a straightforward tuning of the algorithm for particular network environments. In~\cite{Li2014}, the authors use dynamic programming to solve a \ac{NUM} problem for a finite time horizon, for which a bandwidth estimation is computed based on an \ac{EWMA} of the recent segment downloads. The proposed adaptation algorithm is evaluated using buffer size limits between 30 and 50 seconds. A heuristic adaptation algorithm is proposed in~\cite{Le2013} that tries to satisfy constraints on future buffer levels over a finite time horizon using throughput predictions obtained using a modified \ac{EWMA} model where the weights are dynamically adjusted based on the most recent relative prediction error. The algorithm is evaluated against two baseline approaches using a single throughput trace and a buffer size of 20 seconds. The evaluation reveals that the average video quality offered by the proposed algorithm is by 3\% (17\%) higher, the average quality transition magnitude is by 27\% (9\%) lower, and the number of quality transitions by 80\% higher (46\% lower) as compared to the two baseline approaches.


Since having accurate throughput predictions is beneficial for a number of applications, there are dedicated studies that address this subject, see~\cite{He2007} for an overview. Since, however, the main application of TCP was for a long time bulk data transfer, many of those studies predict throughput averaged over much longer time intervals than required for low-delay streaming~\cite{Padhye2000,He2007}. The authors in~\cite{He2007} observed that time series prediction methods perform quite well on the time scale of 50 seconds (\ac{RMSRE} is less than 0.4 for about 90\% of studied traces), but the accuracy strongly varies across studied network paths. Evaluations are often limited to wired networks which, however, are not subject to channel state dynamics and effects caused by the data link layer to the extent seen with wireless technologies~\cite{Padhye2000,He2007,Mirza2010}. Some of the developed approaches use information available only at the sender, or they require cross-layer information, or additional path measurements that need to be supported by the other end-point~\cite{Padhye2000,Mirza2010}. In addition, some analytical models target specific TCP flavors making their performance uncertain given recently developed variants~\cite{Padhye2000}.

Identifying performance goals for adaptive video streaming and expressing them in a way that facilitates objective measurements is an extremely challenging task. It must take into account human perception and cognitive processing --- phenomena influenced by a hard to measure factors. The notion of \acf{QoE} was introduced in an effort to assess these phenomena and to help make them accessible to an objective evaluation process~\cite{ITUT_P10G100_Am2}. The number of factors influencing \ac{QoE} is immense, and many of them have a high level of subjectivity that results in extremely complex modeling~\cite{Reiter2014}.
\ac{QoE} for adaptive video streaming is an important and a fast developing research area~\cite{Balachandran2012,Seufert2014,Reiter2014,Song2014}. Rebuffering, initial delay, and quality fluctuations are factors that have not been part of traditional \ac{QoE} metrics for video, but that have a tremendous impact on user's perception of adaptive video streamed over a best-effort network such as the Internet. In particular, many studies suggest that the number and duration of playback interruptions have the most severe impact on \ac{QoE}~\cite{Conviva2014}. Users are willing to accept a higher initial delay and higher video distortion, if it helps minimize playback interruptions~\cite{Seufert2014,Hossfeld2012a,Quan2008,Singh2012a}. On the other hand, it was observed that video quality fluctuations resulting from dynamically changing the representation can have a negative impact on \ac{QoE}~\cite{Lewcio2011,Yitong2013}. In particular, some studies conclude that a lower average video quality might be tolerated if it helps reduce the amount of representation transitions~\cite{Pessemier2013}. User engagement is another important metric, which is especially of interest for content providers because it directly relates to advertising-based revenue schemes~\cite{Balachandran2013,Conviva2014}.

\chapter{Considered System and its Model}
\label{sec:system_model}

\begin{figure}[t]
\centering
\includegraphics[scale=0.17]{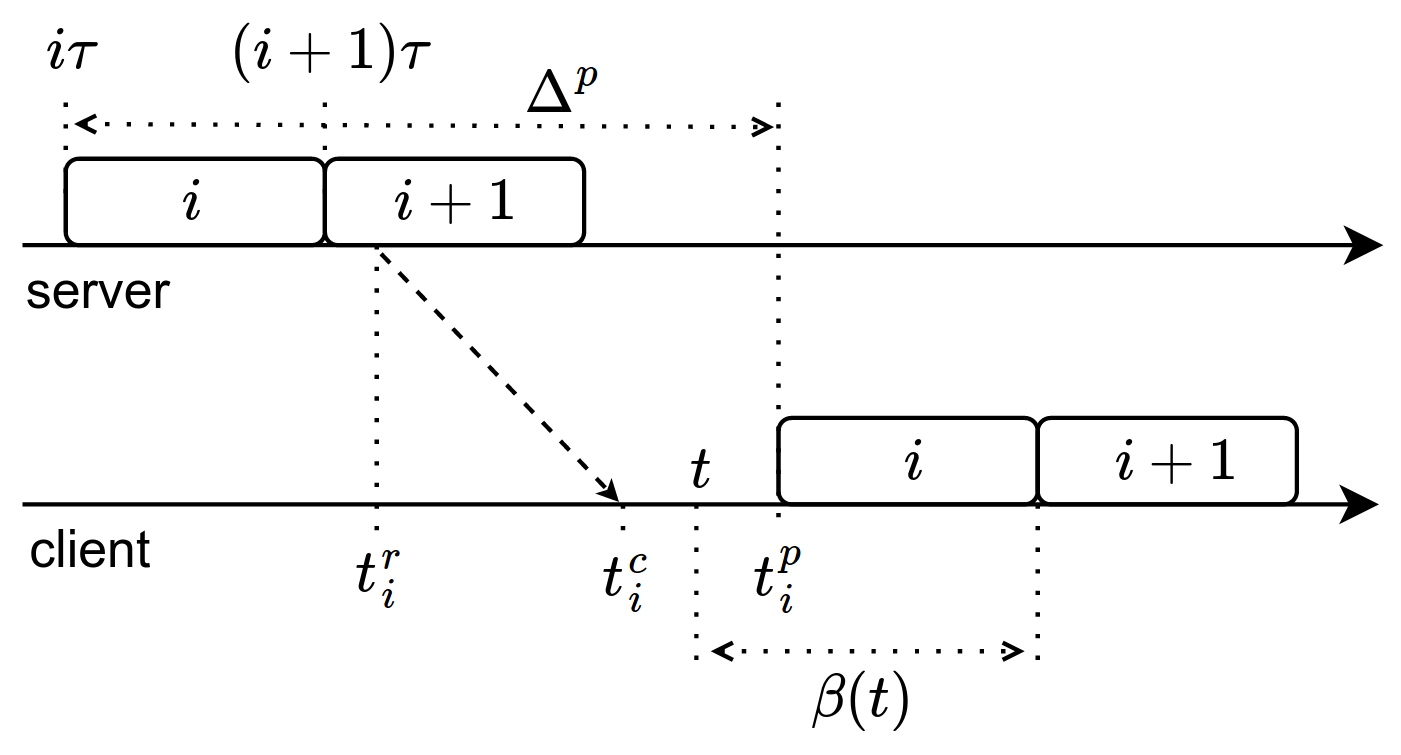}
\caption{Illustration of the used notation}
\label{fig:system_model}
\end{figure}


In a live streaming system, the video content is recorded and published while streaming, in contrast to being prerecorded and stored at the server as in the case of \ac{VoD}. The difference between the time when the content is recorded and the time when it is rendered on the user's device is often termed live latency. In order to provide the ``live'' experience, it is typically constrained by an upper bound. This severely limits the capability of the client to prefetch content to alleviate transport latency variations caused by varying network conditions, thus making the design of the system more challenging. The live latency consists of several components: sever-side processing (cutting, encoding, etc.), publishing (making available for download, distributing among \ac{CDN} nodes, etc.), transport latency (downloading the content), and client-side processing (demultiplexing, decoding, rendering).

In an \ac{HAS} system, the video content is encoded in several representations varying w.r.t. their media characteristics such as spatial resolution, frame rate, compression rate, etc. They are configured by the service provider during the planning phase~\cite{Seufert2014}. Each representation is split into segments, typically containing several seconds of video data, such that switching the representation is feasible on each segment boundary. The client issues \ac{HTTP} requests to download the segments in chronological order, selecting the representation for each of them. After the segment is downloaded, it is stored in the playback buffer until its playback deadline is reached. With live streaming, a segment becomes available for download during the course of the streaming session. If the download is not completed before the playback deadline, the playback is skipped. Since different representations typically have different media bit rates, the client is able to satisfy the latency constraint by dynamically selecting an appropriate representation for each segment. Note that the segment duration affects the client's responsiveness to throughput changes and thus facilitates achieving low latencies. At the same time, however, small segments increase the overhead due to the higher number of \ac{HTTP} requests as well as reduce the video compression efficiency due to the decreased \ac{GOP} size. Typical segment durations lie between 2 and 15 seconds.

Since \ac{HTTP} offers no means to cancel an ongoing request, the only way to prevent wasting bandwidth by downloading the remaining bytes of a segment whose playback has to be skipped is to shutdown the TCP connection. Since opening a new TCP connection is associated with communication overhead, we assume that the client maintains multiple TCP connections, using them in a Round Robin manner in order to keep their internal state such as congestion window size and \ac{RTT} estimation up-to-date. Whenever a TCP connection is closed, other connections are used, while the closed connection is replaced by a new one.

In addition to satisfying latency requirements, one of the main adaptation goals is to maximize \ac{QoE}. In the following, we define \ac{QoE} as the triplet (number of skipped segments, number of quality transitions, average video quality). We use the term video quality to refer to the video distortion, which is typically a concave function of the video bit rate~\cite{Sullivan1998}. As stated previously, it is necessary to consider these factors jointly since optimizing any one parameter individually leads to poor \ac{QoE}. Our approach is to heuristically maximize the average video quality as a function of the pair: number of skipped segments and number of quality transitions, which we define as an operating point. Since the duration of the streaming session is not known in advance (the user might quit the session prematurely), both values are expressed in relative terms: fraction of skipped segments and fraction of segments that result in a transition. The operating point may be defined by the user, the operating system, the client software, or the content provider. It might depend on various factors, such as the nature of the video, the user context, the provider's business model, etc.

In the following, we will introduce the notation used throughout the paper. All time-related variables are real-valued and represent continuous time. The start time of the recorded content is $t=0$. The duration of video content contained in one segment is constant and denoted by $\tau$. We use index $i\in\{0,1,\ldots,n-1\}$ to indicate a particular segment in a stream.  
Segment $i$ contains video material covering the time period $\left[i\tau,\,(i+1)\tau\right]$ and becomes available for download at time $(i+1)\tau$.


We denote the set of available video representations by $\mathcal{R}$, indexed by $j\in\left\{0,1,\ldots,\lvert\mathcal{R}\rvert-1\right\}$. We denote the size of a segment in bits by $s_{ij}$, and by $\bar{r}_{ij}=s_{ij}/\tau$, the \ac{MMBR} of segment $i$ from representation $j$. We denote by $\bar{r}_j=1/n\sum_{i=1}^n{\bar{r}_{ij}}$ the \ac{MMBR} of representation $j$.
We denote the size of a downloaded segment $i$ by $s_i$, from the representation that was used to download it. Consequently, $\bar{r}_i=s_i/\tau$ denotes the \ac{MMBR} of a downloaded segment $i$.

The time when the request to download segment $i$ is sent by the user is denoted by $t^r_i$. Note that it arises from the maximum of two values: the time when the client finished downloading segment $i-1$, and the time when segment $i$ becomes available at the server. $t_i^c$ denotes the time when the last bit of segment $i$ is received by the user. 
We denote the upper bound on the live latency by $\Delta^p$. Consequently, the playback deadline of segment $i$ is $t_i^p=i\tau+\Delta^p$.
The value of $\Delta^p$ bounds the maximum transport latency, which is given by $\Delta^p-\tau$ if other latency components are neglected. Note that the maximum transport latency for individual segments can be smaller since it depends on the playback buffer level at the time of the request.
The playback buffer level at time $t$ is the time until the playback deadline of the next segment whose download is not completed yet: $\beta(t)=\max\left\{t_i^p\;|\;t_i^c\leq t\right\}+\tau-t$. The maximum transport latency for segment $i$ is thus given by $0<\beta\left(t_i^r\right)\leq\Delta^p-\tau$. See Figure~\ref{fig:system_model} for an illustration.

We denote the fraction of segments skipped until time $t$ by $\Sigma(t)\in[0,1]$. When segment $i$ is downloaded and played in representation $j$ different from the representation of the previous successfully downloaded segment, a quality transition occurs. The fraction of segments that were successfully downloaded in a different quality than their predecessors until time $t$ is denoted by $\Omega(t)\in[0,1]$.




Note that the traffic generated by a live streaming client might contain inter-request periods during which the client is waiting for the next segment to become available. When computing average application layer throughput, we exclude the inter-request periods in order to obtain an estimate of the throughput that was actually achieved during data transmission.
We first compute the segment throughput for each downloaded segment $i$ as $\rho_i=s_i/\left(t^c_i-t^r_i\right)$. Note that this computation accounts for the round-trip delay. We then compute the average application layer throughput for the time interval $\left[t_1,t_2\right]$ as
\begin{equation}
\label{eq:thrpt}
\rho\left(t_1,t_2\right)=\frac{\sum_{i=l_1}^{l_2}s_i\cdot\lvert\left[t_i^r,t_i^c\right]\cap\left[t_1,t_2\right]\rvert}{\sum_{i=l_1}^{l_2}\lvert\left[t_i^r,t_i^c\right]\cap\left[t_1,t_2\right]\rvert}\,,
\end{equation}
where $l_1$ corresponds to the last segment requested before $t_1$, $l_2$ is the first segment whose download was completed after $t_2$, and $\lvert[a,b]\rvert=b-a$. For incomplete downloads, $t_i^c$ must be replaced by the time when the download was interrupted, while $s_i$ must be replaced by the number of actually downloaded bytes. For time intervals, for which the denominator equals 0, $\rho\left(t_1,t_2\right)$ is not defined.



\chapter{\acs{LOLYPOP} --- Adaptation Algorithm for Low-Delay Live Streaming}
\label{sec:adaptation}






In this chapter, we present our design of \ac{LOLYPOP}, a novel prediction-based adaptation algorithm for low-delay streaming over \ac{HTTP}.

\section{Algorithm description}
\label{sec:algo}

As described in Chapter~\ref{sec:system_model}, the goal of \ac{LOLYPOP} is to maximize the average video quality as a function of the operating point, defined by the pair $\left(\Sigma,\Omega\right)$. The input parameters controlling the reached operating point are $\Sigma^*\in[0,1]$, which controls the fraction of skipped segments (we will describe it in more details in the following), and $\Omega^*\in[0,1]$, which is an upper bound on the (relative) number of quality transitions. The output of the algorithm is the representation for the next segment to be downloaded. The approach leverages throughput predictions and prediction error estimations to compute the probability $P^p_{ij}$ that the download of segment $i$ in quality $j$ will be completed before its playback deadline. Computation of $P^p_{ij}$ is described in detail in Section~\ref{sec:ur_prob}. For now, we assume that $P^p_{ij}$ is given.

Let us consider the decision about downloading the segment $i$. First, \ac{LOLYPOP} identifies the highest representation $j'$ such that the probability for missing the playback deadline is bounded by $\Sigma^*$, i.e. $1-P^p_{ij'}\leq\Sigma^*$. If no representation satisfies this condition or the download success probabilities cannot be computed (e.g., because the streaming session has just started or after a period of zero throughput), $j'$ is set to 0.

In the second step, \ac{LOLYPOP} computes $\Omega(t)$, the fraction of segments that were played in a different quality than their predecessor. If $\Omega(t)>\Omega^*$, and $j'>j^\leftarrow$, where $j^\leftarrow$ is the representation of the last successfully downloaded segment $i^\leftarrow < i$, representation $j^\leftarrow$ is selected in order to prevent $\Omega(t)$ from further exceeding the upper bound $\Omega^*$.
The pseudocode for the described algorithm is presented in Figure~\ref{alg:lolypop}.

\begin{figure}
\begin{algorithmic}[1]
\Input{}$t_i^r,\;t_i^p,\;\Sigma^*,\;\Omega^*$ \Comment{\parbox[t]{0.61\linewidth}{Invocation time, playback deadline, config.\ parameters}}
\Input{}$\Omega\left(t_i^r\right)$ \Comment{\parbox[t]{.61\linewidth}{Current value for the relative number of quality transitions}}
\Input{}$\left(P^p_{ij},\,j\in\left\{0,\ldots,\lvert\mathcal{R}\rvert-1\right\}\right)$ \Comment{\parbox[t]{.53\linewidth}{Estimated download success probabilities, or -1}}
\Input{}$j^{\leftarrow}\in\left\{0,\ldots,\lvert\mathcal{R}\rvert-1\right\}$ \Comment{\parbox[t]{.53\linewidth}{Repr.\ of the last successfully downloaded segment}}
\Output{}$j^*$ \Comment{\parbox[t]{0.42\linewidth}{Selected representation}}
\Require{}$(i+1)\tau\leq t_i^r<t_i^p\leq t_i^r+T_{\text{max}}$ \Comment{\parbox[t]{.42\linewidth}{Segment $i$ available, playback deadline not passed and within prediction horizon}}
\If{$P^p_{ij}=-1,\;\forall j\in\left\{0,\ldots,\lvert\mathcal{R}\rvert-1\right\}$} \Comment{\parbox[t]{.42\linewidth}{No estimation available}}
  \State{}$j^*=0$ \Comment{\parbox[t]{.42\linewidth}{Select lowest representation}}
\Else{}\Comment{\parbox[t]{.61\linewidth}{An estimation of download success probabilities is available}}
  \State{}$j'=\max\left(\left\{0\right\}\cup\left\{j\in\left\{0,\ldots,\lvert\mathcal{R}\rvert-1\right\}\,|\,1-P^p_{ij}\leq\Sigma^*\right\}\right)$ \Comment{\parbox[t]{.27\linewidth}{Max.\ representation satisfying $\Sigma^*$, 0 if none}}
  \If{$\Omega\left(t_i^r\right)\leq\Omega^*$} \Comment{\parbox[t]{.53\linewidth}{Transition to a higher representation is possible}}
    \State{}$j^*=j'$
  \Else{} \Comment{\parbox[t]{.53\linewidth}{Transition to a higher representation is not possible}}
    \State{}$j^*=\min\left(j',j^{\leftarrow}\right)$
  \EndIf{}
\EndIf{}
\end{algorithmic}
\caption{Pseudocode of \ac{LOLYPOP}}
\label{alg:lolypop}
\end{figure}

The intuition for letting \ac{LOLYPOP} switch to a lower representation, even if the upper bound on the quality transitions is exceeded, is that preventing a quality decrease can significantly increase the number of skipped segments. According to a large-scale study of user engagement (time before the user quits a streaming session), it is always better to drop video quality than to let the streaming stall~\cite{Conviva2014}.

\section{Tuning into the stream}\label{sec:tune_in}

When the client is about to start a new streaming session, it has to decide which segment to download first and in which representation. The client might start with the newest segment among those available for download, maximizing the probability that the first segment will be downloaded before its playback deadline but increasing the initial delay. In contrast, taking the oldest segment whose playback deadline is sufficiently far into the future minimizes the initial delay, given that the download can be completed in time. \ac{LOLYPOP} adopts the latter approach and selects the first segment $i_0$ as the oldest segment whose playback deadline is at least $\tau$ seconds into the future: $i_0=\min\left\{i\geq 0\;|\; (i+1)\tau\leq t \wedge t_i^p \geq t + \tau \right\}$, where $t$ is the time when a user tunes into the stream. Furthermore, it downloads the first segment in the lowest quality in order to minimize the risk of missing its playback deadline and thus unnecessarily increasing the initial delay. Due to the small segment duration, the low quality of the first segment has negligible impact on the average video quality of the streaming session. The intuition is that the available network resources should at least support the download of a segment in its lowest quality in less time than the segment duration. If the first segment can be downloaded before its playback deadline, as expected, the start-up delay will thus lie in the interval $\left[\tau,\,\Delta^p-\tau\right]$, which can be seen by transforming the equation for $i_0$, using $t_i^p=i\tau+\Delta^p$. \ac{LOLYPOP} applies the same decision process when the client skips a segment and has to select a segment to proceed with.

\chapter{\acs{TCP} Throughput Traces}
\label{sec:traces}

As previously stated, \ac{LOLYPOP} leverages estimations of probabilities $P_{ij}^p$ that segment $i$ can be downloaded in representation $j$ before its playback deadline. In order to develop an efficient estimator, it requires a data set, that is, a collection of \ac{TCP} throughput traces from IEEE 802.11 \acp{WLAN}, to evaluate the accuracy and error distributions of different time-series prediction methods. In addition, such a data set is required to evaluate the proposed adaptation algorithm under different network conditions. Due to the targeted transport latency of a few seconds, the required data set must contain throughput averages computed over relatively small time intervals: 1 second or less. To the best of our knowledge, there exists no such publicly available data set. Therefore, we collected a representative set of \ac{TCP} throughput traces in IEEE 802.11 networks in different environments at various locations throughout Berlin, Germany and Irbid, Jordan. Our selected locations include public hotspots (indoor and outdoor), campus hotspots, and access points in residential environments. The traces were collected using laptops running Ubuntu 13.04 and Ubuntu 14.04 operating systems with default \ac{MAC} and \ac{TCP} configurations.

We collected 127 traces of continuous downstream \ac{TCP} flows, lasting between 600 and 3600 seconds each. In order to focus on the more challenging scenarios, we removed traces with a \acf{CV} of less than 0.1 resulting in 92 traces with a total length of 45 hours. As a sender, we used either a server located at the TU Berlin campus (running Ubuntu 12.04) or an Amazon EC2 micro instance located in Ireland (running Ubuntu 14.04). More information about the traces, as well as an example figure depicting the throughput of one complete trace, are provided in Appendix~\ref{app:traces}. All traces are available upon request.

\begin{figure}
\centering
\includegraphics[scale=1.0]{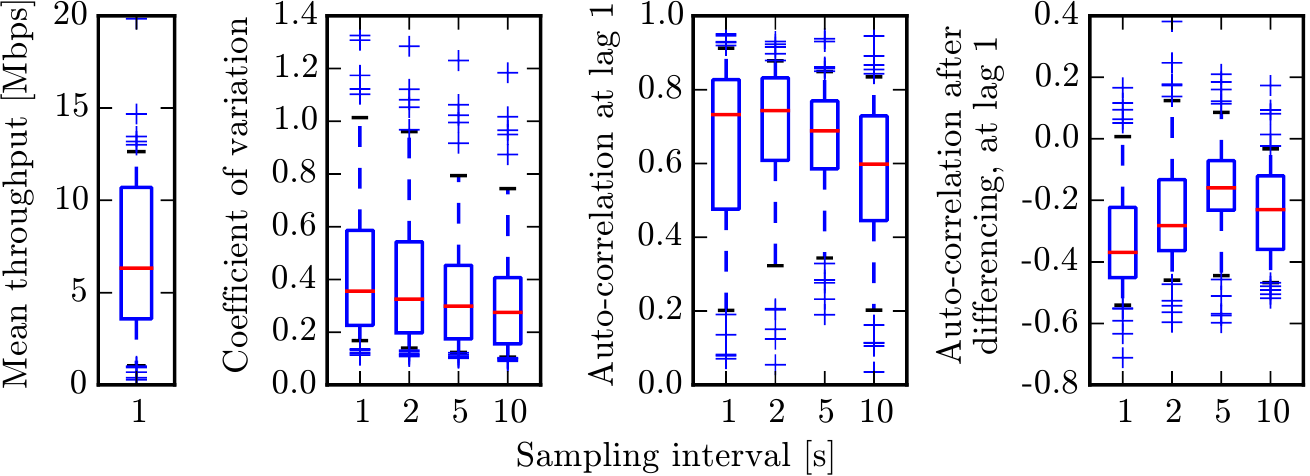}
\caption{Trace statistics: mean throughput (equal for all sampling intervals), \acf{CV}, auto-correlation at lag 1, and auto-correlation, after differencing, at lag 1. Horizontal line: median, box: quartiles, whiskers: 0.5 and 0.95 quantiles, flier points: outliers.}
\label{fig:trace_statistics}
\end{figure}

Figure~\ref{fig:trace_statistics} provides an illustration of throughput variability and temporal correlation that are among the main factors affecting predictability. The figure shows boxplots for selected sampling intervals of the mean throughput, \acf{CV}, auto-correlation at lag 1, and auto-correlation after differencing at lag 1.
The \ac{CV} is defined as the standard deviation divided by the mean. Auto-correlation at lag 1 shows how probable it is that a large throughput value is followed by another large value (auto-correlation close to 1) or a small value (auto-correlation close to -1). A value close to 0 indicates no temporal correlation between subsequent values. Auto-correlation after differencing quantifies the correlation of throughput changes.
The computed statistics confirm that our traces cover a broad range of network conditions. 50\% of the traces have a mean throughput between 4 and 11 Mbps, while for 90\% of traces the mean lies between approximately 1 and 13 Mbps. The range of throughput fluctuations, represented by the \ac{CV}, varies approximately from 0.1 to 1.3.
An interesting observation is that 75\% of traces show auto-correlation values of over 0.6 at a sampling interval of 2 seconds, while 95\% still have auto-correlation values over 0.3, indicating significant temporal correlation between subsequent measurements. At the same time, the time series of throughput changes exhibits a strong negative auto-correlation, indicating that a throughput increase is likely to be followed by a throughput decrease.


\chapter{Short-Term TCP Throughput Prediction}
\label{sec:prediction}

In this chapter, we present our approach to estimating download success probabilities required by \ac{LOLYPOP}. It is based on predicting \ac{TCP} throughput and estimating the relative prediction error distribution.


\section{Methodology}
\label{sec:prediction_methodology}

Our goal is to estimate the probabilities $P_{ij}^p$ that the download of $s_{ij}$ bytes, requested at time $t_i^r$, will be completed by the time $t_i^p$. We achieve this by using a time series prediction complemented by estimating the relative prediction error distribution.
A time series prediction method uses several past values to compute one or several future values.
Thus, from $\rho\left(t-iT, t-(i-1)T\right)$, $i\in\left\{1,\ldots,n\right\}$, it computes predictions $\hat{\rho}\left(t+(i-1)T,t+iT\right)$, $i\in\left\{1,\ldots,k\right\}$, where $T$ is the averaging interval.

As described in Chapter~\ref{sec:system_model}, the upper bound on the download duration $t_i^p-t_i^r$ for segment $i$ takes values from the range $\left(0, \Delta^p-\tau\right]$, depending on the completion time of the preceding download. We, therefore, have the following two options to compute predictions. We can fix $T$ and whenever $t_i^p-t_i^r>T$, compute a prediction for multiple steps into the future or, we compute predictions using multiple values for $T$ and then use the smallest one such that $T\geq t_i^p-t_i^r$. In the course of the study, we observed that the latter approach performs significantly better. Consequently, we focus on predictions on multiple time scales, always for one step into the future. We focus on time scales from 1 to 10 seconds because of their relevance to low-delay streaming.

Given a prediction $\hat{\rho}\left(t_1,\,t_2\right)$ and the corresponding measured throughput $\rho\left(t_1,\,t_2\right)$, we compute the relative prediction error as
\begin{equation}\label{eq:epsilon}
\epsilon\left(t_1,\,t_2\right)=\\
\frac{\lvert\max{\left(\hat{\rho}\left(t_1,\,t_2\right),\,\rho_{\min}\right)}-\max{\left(\rho\left(t_1,\,t_2\right),\,\rho_{\min}\right)}\rvert}{\max{\left(\rho\left(t_1,\,t_2\right),\,\rho_{\min}\right)}}\,.
\end{equation}
Here, the maximum operator prevents a distortion of results whenever $\rho\approx 0$ or $\hat{\rho}\leq 0$. In the following, we set $\rho_{\min}=10$ kbps. We separately evaluate the overestimation and the underestimation errors, due to their different error ranges ($(0,\infty)$ and $(0, 1]$ respectively) and due to their different impacts on the adaptation. An overestimation increases the risk of skipping a segment, which has the strongest impact on \ac{QoE}. An underestimation decreases the risk of interruptions but reduces the video quality.

\section{Prediction methods}\label{sec:methods}




We evaluated a number of time series prediction methods, including \ac{SMA}, linear extrapolation, \ac{CSS}, several flavors of exponential and double exponential smoothing, \ac{ARIMA}, and several machine learning based methods~\cite{Chatfield2013}.
Our throughput prediction results will focus on three simple methods: \ac{SMA}, linear extrapolation, and double exponential smoothing (Holt-Winters). A brief description of these methods is provided in Appendix~\ref{app:prediction_methods}.
We abbreviate the methods by $\langle$type$\rangle$:$\langle n\rangle$:$\langle$parameters$\rangle$, where $\langle$type$\rangle$ is the name of the method, $n$ is the number of past throughput values used as input, and $\langle$parameters$\rangle$ include further optional configuration parameters. For example, SMA:$n$:ar denotes \ac{SMA} with arithmetic mean, and HW:$n$:mse denotes Holt-Winters with \ac{MSE} used for parameter tuning.


\begin{figure}
\centering
\includegraphics[scale=1]{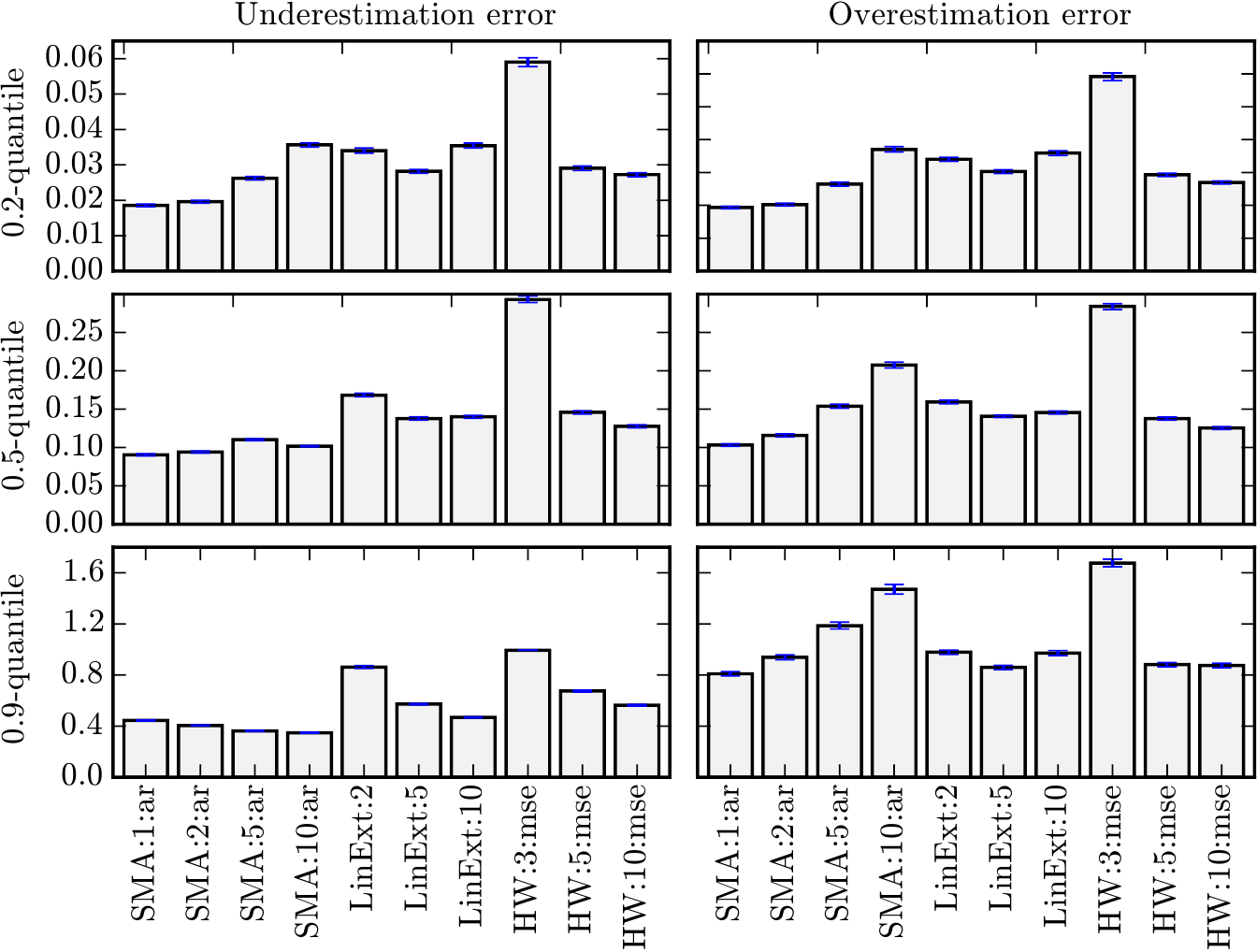}
\caption{Relative prediction error quantiles for the complete data set, on the time scale of 5 seconds. Left column shows the underestimation error, right column the overestimation error. The error bars show confidence intervals for the confidence level of 0.95. See Section~\ref{sec:performance} for details.}
\label{fig:err_rel_cdf_all_traces_as_one}
\end{figure}

\section{Evaluation of the prediction accuracy}\label{sec:performance}


We evaluate the prediction methods in two steps. First, we compare the relative prediction errors over the joint data set from all of the traces to identify the method that performs best over a broad range of network environments. In the second step, we evaluate how the prediction accuracy varies over individual traces.

To compare the prediction accuracy over the complete data set, we computed the 0.2-quantiles, 0.5-quantiles, and 0.9-quantiles of the prediction error. For example, a 0.2-quantile of 0.3 means that in 20\% of all collected data points, the relative prediction error is below 0.3. Another example: a 0.9-quantile of 0.8 means that less than 10\% of data points have an error over 0.8. The results for the sampling interval of 5 seconds are shown in Figure~\ref{fig:err_rel_cdf_all_traces_as_one}. We observe that \ac{SMA}:1:ar has the best performance except for the 0.9-quantile of the underestimation error, where SMA:10:ar is the best performing method. For 50\% of the data points, \ac{SMA}:1:ar results in an overestimation error that does not exceed 10\%, while only less than 10\% of data points have an overestimation error of 80\% and larger. This is somewhat surprising since \ac{SMA}:1:ar is the most na\"{\i}ve method that uses only the most recent measurement as prediction. It also has a much lower computational complexity than methods such as Holt-Winters due to the optimizations involved in tuning the configuration parameters of the latter for every new prediction. It seems that taking into account the trend in the past measurements does not improve the prediction quality. This is consistent with the observation that in many traces the differences in subsequent throughput measurements show a negative correlation, as depicted in Figure~\ref{fig:trace_statistics}. Also, methods using a small number of history points implicitly detect level shifts and do not propagate outliers. These two issues were reported to be known challenges in TCP throughput prediction~\cite{He2007}.

\begin{figure}
\centering
\includegraphics{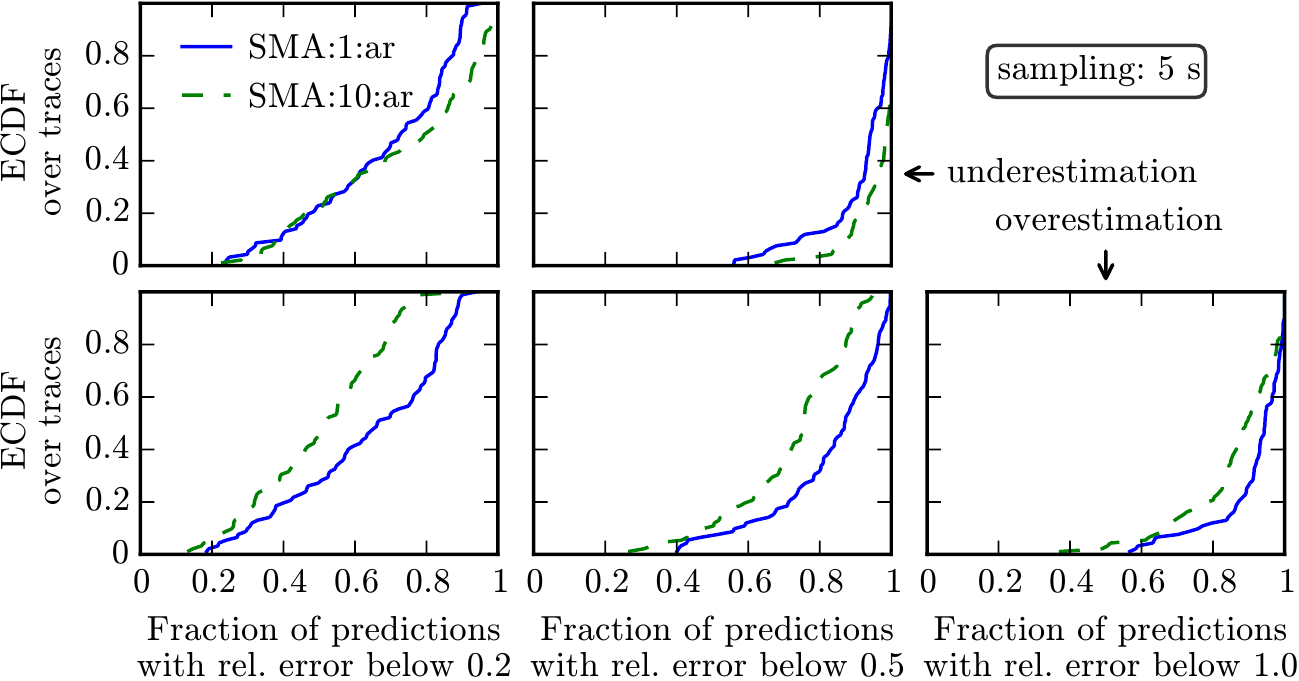}
\caption{Performance of SMA:1:ar and SMA:10:ar for individual traces. See Section~\ref{sec:performance} for details.}
\label{fig:compare_quantiles}
\end{figure}

In the second step, we evaluate how the prediction accuracy varies over the individual traces. For each trace and method, we compute the fractions of predictions with a relative error less than 0.2, 0.5, and 1.0. The \acp{ECDF} of these fractions over individual traces for the sampling interval of 5 seconds, is shown in Figure~\ref{fig:compare_quantiles}. The first/second/third column shows for each trace the fraction of measurements with a relative error below 0.2/0.3/0.5 respectively.
For example, the point $(0.8, 0.3)$ on the solid line in the middle column, bottom row, represents a trace, where 80\% of the overestimations have a relative error of 0.5 or less. Points below (y-value less than 0.3) correspond to traces that have worse performance, while points above (y-value over 0.3) correspond to traces with better performance.
The fact that the graph passes through point $(0.8, 0.3)$ means that in 70\% of traces (sampled at 5 s), less than 20\% of the overestimations have a relative error of 0.5 or more.

From Figure~\ref{fig:compare_quantiles}, we observe that the prediction strongly varies across traces. There are traces, where 90\% of overestimations have an error less than 20\%, while 100\% of predictions have an error less than 50\%. A video client might account for a relative error of this magnitude by using a fixed safety margin, that is, by always selecting a media bit rate which is 20\% smaller than the predicted throughput. There are, however, traces where almost 60\% of the overestimations have a relative error of greater than 50\%, while more than 40\% of the overestimations still have an error greater than 100\%. Setting a high fixed safety margin to account for such ``bad'' traces would result in significant underutilization of network resources, lower media bit rate and thus lower \ac{QoE} in the ``well-behaving'' traces. On the other hand, selecting a low fixed safety margin would increase the total number of skipped segments in the ``bad'' traces. Consequently, we have to complement a time series prediction with an estimation of the prediction error distribution.


Finally, in all of the studied traces, we observed that the probability for occurrences of underestimations and overestimations are well balanced on all time scales. Both occur in approximately $50\%\pm 5\%$ of cases. At the same time, they exhibit a significant temporal correlation: the probability that an underestimation is followed by an overestimation and vice versa is significantly over 50\% for most traces for all time scales, exceeding 80\% or even 90\% in some cases. This observation is directly related to the negative correlation of the throughput after differencing, depicted in Figure~\ref{fig:trace_statistics}. The distribution of per-trace values is depicted in Appendix~\ref{app:correlation_errors}.


\section{Estimating the relative prediction error}\label{sec:modeling}

In order to estimate the download success probability, it is not sufficient to perform a time series prediction because the uncertainty of such a prediction can be quite high (as shown previously) and because it can vary across different network environments. Although there are approaches that allow to explicitly predict a distribution such as Gaussian process method~\cite{Rasmussen2006}, approaches such as \ac{SMA} do not have this capability. Therefore, we complement the predicted value by an estimate of the relative prediction error distribution.
A straightforward approach is to use the \ac{ECDF} of past prediction errors, and to account for the long-term non-stationarities by discarding values whose age exceeds a certain threshold. Another approach is to select a distribution type and to fit its parameters dynamically from the data. In our evaluation, we will use the former method, since it results in good performance and since with the latter method the computation of the model parameters involves an optimization step, which is resource-consuming. In Appendix~\ref{app:fitting_distributions} we present our results on fitting several well-known distribution types to relative prediction errors: exponential, normal, logistic, and Lomax (shifted Pareto)~\cite{Johnson1994}. We observe that the Lomax distribution provides the best fit. We therefore recommend to use the Lomax distribution to model prediction errors when evaluating adaptation approaches with synthetic data, as done in~\cite{Yin2014}, for example.





\section{Estimating download success probabilities}
\label{sec:ur_prob}

In this section, we describe our approach to estimating download success probabilities $\big(P^p_{ij}, j\in\left\{0,\ldots,\lvert\mathcal{R}\rvert-1\right\}\big)$ that at time $t^r_i$ $s_{ij}$ bytes can be downloaded in representation $j$ before its playback deadline $t^p_i$.
We denote by $T_{\text{max}}\in\mathbb{N}$ the maximum prediction horizon in seconds. Consequently, at time $t\in\mathbb{N}$, the client computes the average application layer throughput (defined in~\eqref{eq:thrpt}) for time intervals $\left[t-T,t\right]$ for $T\in\left\{1,\ldots,T_{\text{max}}\right\}$, followed by computing throughput predictions for time intervals $\left[t,t+T\right]$.
If the throughput cannot be computed, no prediction is provided either.
Finally, the client computes the relative prediction error for the interval $\left[t-T,t\right]$ as
\begin{equation}
\label{eq:epsilon_sign}
\tilde{\epsilon}\left(t-T,t\right)=\frac{\max{\left(\hat{\rho}\left(t-T,t\right),\rho_{\min}\right)}-\max{(\rho\left(t-T,t\right),\rho_{\min})}}{\max{(\rho\left(t-T,t\right), \rho_{\min})}}\,.
\end{equation}
In contrast to the definition in~\eqref{eq:epsilon}, $\tilde{\epsilon}\left(t-T,t\right)\in\left(-1,\infty\right)$ is defined without taking the absolute value for the purposes of presentation.

We assume that predictions are computed every second, so that at time $t_i^r$, the most recent predictions were computed at time $\lfloor t_i^r\rfloor$. In order to calculate the download success probabilities, the client determines the smallest time interval containing $\left[t_i^r,t_i^p\right]$ for which a prediction is available. Note that it is not necessarily $\left[\lfloor t_i^r\rfloor,\lceil t_i^p\rceil\right]$ since, due to the distribution of inter-request delays or due to a throughput outage, a prediction for this time interval might not be available. Let $t_i^\pi,\,T^*\in\mathbb{N}$ be determined such that $\left[t_i^\pi,t_i^\pi+T^*\right]$ is the shortest time interval containing $\left[t_i^r,t_i^p\right]$ for which a prediction is available, and let $\hat{\rho}_{i}=\hat{\rho}\left(t_i^\pi,\,t_i^\pi+T\right)$ be the corresponding throughput prediction. $\epsilon_{i}=\epsilon\left(t_i^\pi,\,t_i^\pi+T\right)$ and $\tilde{\epsilon}_{i}=\tilde{\epsilon}\left(t_i^\pi,\,t_i^\pi+T\right)$ shall denote the relative prediction errors, as defined in~\eqref{eq:epsilon} and~\eqref{eq:epsilon_sign}. Further, we denote by $\Phi_{i}^u\left(\epsilon_{i}\right)$ and $\Phi_{i}^o\left(\epsilon_{i}\right)$ the estimated \ac{CDF} of the underestimation and overestimation errors for $\hat{\rho}_{i}$, computed at $t_i^\pi$.
Finally, $P_{i}^u\in[0, 1]$ shall denote the relative frequency of underestimations.
With the introduced notation, the \ac{ECDF} for $\tilde{\epsilon}_{i}$ is given by
\begin{equation}
\Phi_{i}\left(\tilde{\epsilon}_{i}\right)=
\begin{cases}
P_{i}^u \cdot \Phi_{i}^u\left(\epsilon_{i}\right) &\text{for}\;\tilde{\epsilon}_{i}<0\\
P_{i}^u + \left(1-P_{i}^u\right) \cdot \Phi_{i}^o\left(\epsilon_{i}\right) &\text{otherwise}\,.
\end{cases}
\end{equation}
Consequently, the download success probability $P_{ij}^p$ can be estimated as
\begin{equation}
P^p_{ij}=P\left[\frac{s_{ij}}{t_{i}^p-t_i^r}\leq\frac{\hat{\rho}_{i}}{1+\tilde{\epsilon}_{i}}\right]=\Phi_{i}\left(\frac{\hat{\rho}_{i}\left(t_{i}^p-t_i^r\right)}{s_{ij}}-1\right)\,.
\end{equation}

\chapter{Evaluation}
\label{sec:evaluation}


We evaluated the performance of \ac{LOLYPOP} using our collected throughput traces and comparing it against the state-of-the-art algorithm FESTIVE~\cite{Jiang2014} as a baseline. The setting and results are presented in the following.


\section{Evaluation setting}

We implemented both \ac{LOLYPOP} and FESTIVE in a streaming client prototype written in Python\footnote{http://www.python.org}.
We equipped the developed prototype with a feature that allowed it to be executed in virtual time using a throughput trace file as input, thus allowing for a simulative evaluation using collected traces.

We used Big Buck Bunny\footnote{http://www.bigbuckbunny.org} as video content, which is an animated movie of approximately 10 minutes duration. This video was selected due to the availability of raw video data, allowing us to generate representations with high \acp{MMBR}. We encoded 9 representations with \acp{MMBR} distributed between 100 and 20000 kbps with exponentially increasing intervals: 101, 194, 377, 730, 1415, 2743, 5319, 10314, and 20000 kbps, using the H.264/MPEG-4 AVC~\cite{Wiegand2003a} compression format. The chosen intervals correspond to a roughly linear increase of the video quality in terms of \ac{PSNR}~\cite{Sullivan1998}. The encoding was performed using the avconv\footnote{http://libav.org/avconv.html} utility using two passes, with a configuration targeting at low \ac{MMBR} variations among individual segments.

The evaluation was performed using an upper bound on the transport latency of 3 seconds, corresponding to 1.5 times the segment duration. Neglecting segmentation overhead at the server and decoding overhead at the client, this corresponds to an overall live latency of 5 seconds. Each streaming session lasted for 5 minutes.

We evaluated \ac{LOLYPOP} with different values for the configuration parameters $\Sigma^*$ and $\Omega^*$. The goal was to explore the range of operating points with $\Sigma\in\left[0, 0.1\right]$ and $\Omega\in\left[0, 0.5\right]$. We used $\Sigma^*\in\{$0.005, 0.01, 0.02, 0.03, 0.04, 0.05, 0.06, 0.07, 0.1, 0.15, 0.2, 0.25, 0.3, 0.35, 0.4, 0.45, 0.5, 0.55, 0.6, 0.65, 0.7, 0.75, 0.8, 0.85, 0.9, 0.95$\}$ and $\Omega^*\in\{$0.001, 0.005, 0.008, 0.01, 0.02, 0.03, 0.04, 0.05, 0.06, 0.07, 0.08, 0.09, 0.1, 0.15, 0.2, 0.3, 0.5$\}$. In total, we evaluated 442 configurations. Note that we used $\Sigma^*$ values that are much higher than the values for $\Sigma$ we want to achieve. This is due to the observation that tight restrictions on the number of quality transitions $\Omega^*$ results in much lower numbers of skipped segments than the value used for $\Sigma^*$.

The FESTIVE adaptation algorithm was evaluated with a broad range of values around the default configuration in~\cite{Jiang2014}. We vary the values for $\alpha$ (controlling the trade-off between the average quality and the quality fluctuations), $p$ (safety margin between the estimated bandwidth and the selected \ac{MMBR}), and $k$ (controlling the amount of quality fluctuations by enforcing a minimum distance between quality transitions). We used $\alpha\in\{$5, 6, 7, 8, 9, 10, 11, 12, 13, 14, 15, 16, 17, 18, 19, 20$\}$, $p\in\{$0.4, 0.45, 0.5, 0.55, 0.6, 0.65, 0.7, 0.75, 0.8, 0.85, 0.9, 0.95$\}$, and $k\in\{$1, 2, 3, 4, 5, 6, 7, 8, 9, 10, 15, 20, 30, 40, 50$\}$. In total, we evaluated 2880 configurations. We disabled the randomizer feature of FESTIVE since it requires delaying requests. With low-delay streaming, the randomizer feature can lead to an increased number of skipped segments. We did not relax the restriction of FESTIVE that it switches the representation at most one step at a time since we considered it as one of its core features. Finally, we would like to point out that FESTIVE bases its decisions upon the knowledge of the \ac{MMBR} of a representation, while LOLYPOP uses the segment size of the next segment.

\begin{figure}
\centering
\includegraphics[scale=1]{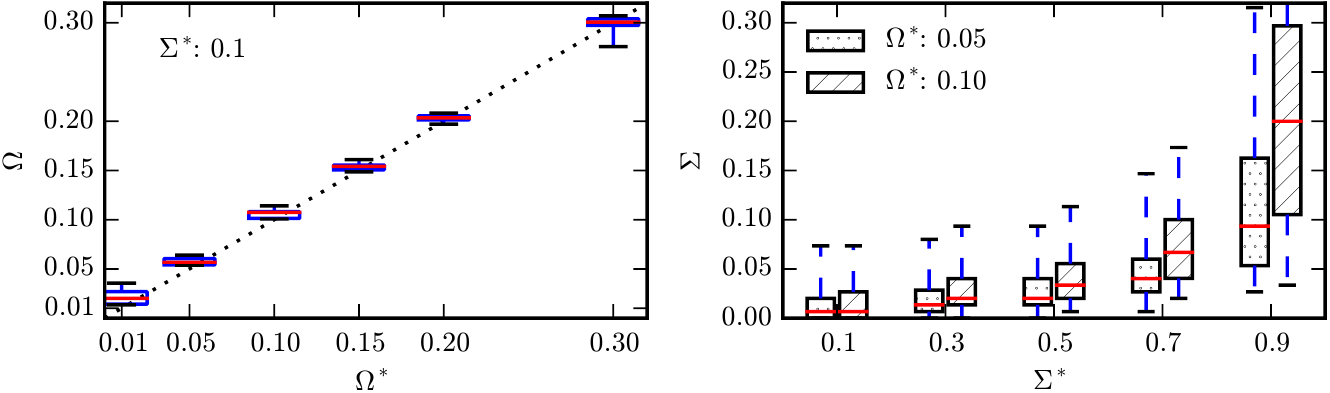}
\caption{$\Omega$ as function of $\Omega^*$ (left), and $\Sigma$ as function of $\Sigma^*$ (right), for \ac{LOLYPOP}. The distributions over the traces are shown as boxplots, where the horizontal line represents the median, the box represents the quartiles, and the whiskers represent the 0.05 and 0.95 quantiles.}
\label{fig:sigma_omega_vs_sigma_omega_star_boxplot}
\end{figure}

\section{Evaluation results}
\label{sec:evaluation_results}

The evaluation goals are to understand the dependency of the reached operating point on the algorithm configuration, to explore the region of reachable operating points, and to evaluate the average video quality as a function of the operating point.

\begin{figure}
\centering
\includegraphics[scale=1]{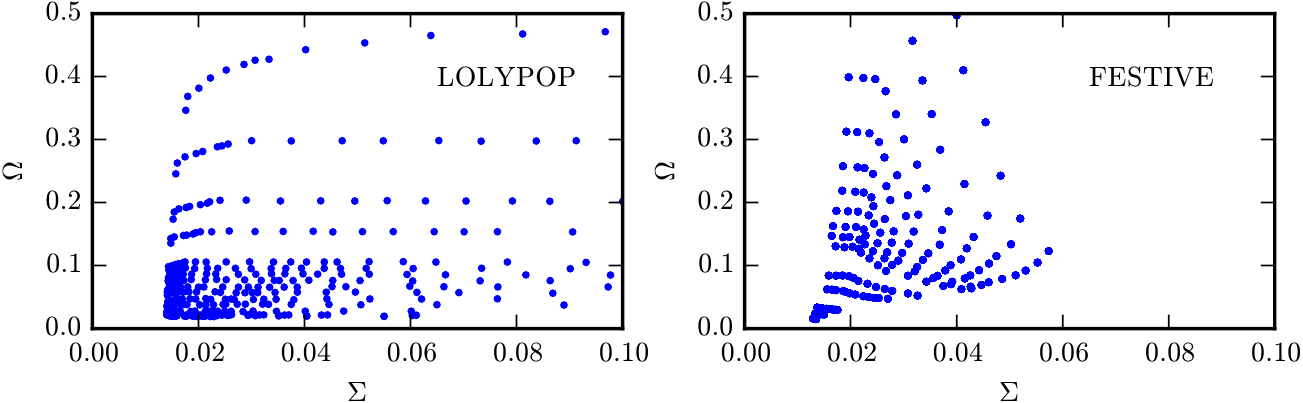}
\caption{Scatter plots of covered $\left(\Sigma,\,\Omega\right)$ regions for \ac{LOLYPOP} (left), and FESTIVE (right).}
\label{fig:sigma_omega_region}
\end{figure}

First, we study the dependency of the reached operating point $\left(\Sigma,\Omega\right)$ on the input parameters $\Sigma^*$ and $\Omega^*$. Figure~\ref{fig:sigma_omega_vs_sigma_omega_star_boxplot} (left) illustrates the ability of \ac{LOLYPOP} to satisfy the upper bound on the number of quality transitions $\Omega^*$, by depicting $\Omega$ as a function of $\Omega^*$ exemplarily for $\Sigma^*=0.1$. The graphs for other values of $\Sigma^*$ are almost identical and are omitted. We observe that \ac{LOLYPOP} is able to enforce the upper bound on the number of quality transitions quite accurately. One reason for the slight overshoot is that we always allow downward quality transitions. Also note that a value of $\Omega=0.01$ means that during the whole streaming session, there are only 3 quality transition, which, in a wireless network, is an extremely low value. Figure~\ref{fig:sigma_omega_vs_sigma_omega_star_boxplot} (right) illustrates the dependency of $\Sigma$ on $\Sigma^*$ for two values of $\Omega^*$: 0.05 and 0.1. We observe that $\Sigma$ is significantly lower than $\Sigma^*$ and that a lower value for $\Omega^*$ decreases $\Sigma$ even further. The intuition behind that is that whenever $\Omega^*$ is exceeded during the course of a streaming session, only downward quality transitions are permitted.

Next, we evaluate the region of reachable operating points. The broader this region the more flexible the algorithm can be tuned to the \ac{QoE} requirements defined for a streaming session.  Figure~\ref{fig:sigma_omega_region} shows scatter plots of achieved $\left(\Sigma,\,\Omega\right)$ values for \ac{LOLYPOP} (left) and FESTIVE (right). We observe that the studied \ac{LOLYPOP} configurations cover a broader range of $\left(\Sigma,\,\Omega\right)$ values and thus enable a more flexible adjustment to user and/or application profiles. Note that a low value of $\Sigma$ and/or $\Omega$ alone is not an indicator of high \ac{QoE} since it might be achieved by selecting an unnecessary low video quality.

\begin{figure}
\centering
\includegraphics[scale=1]{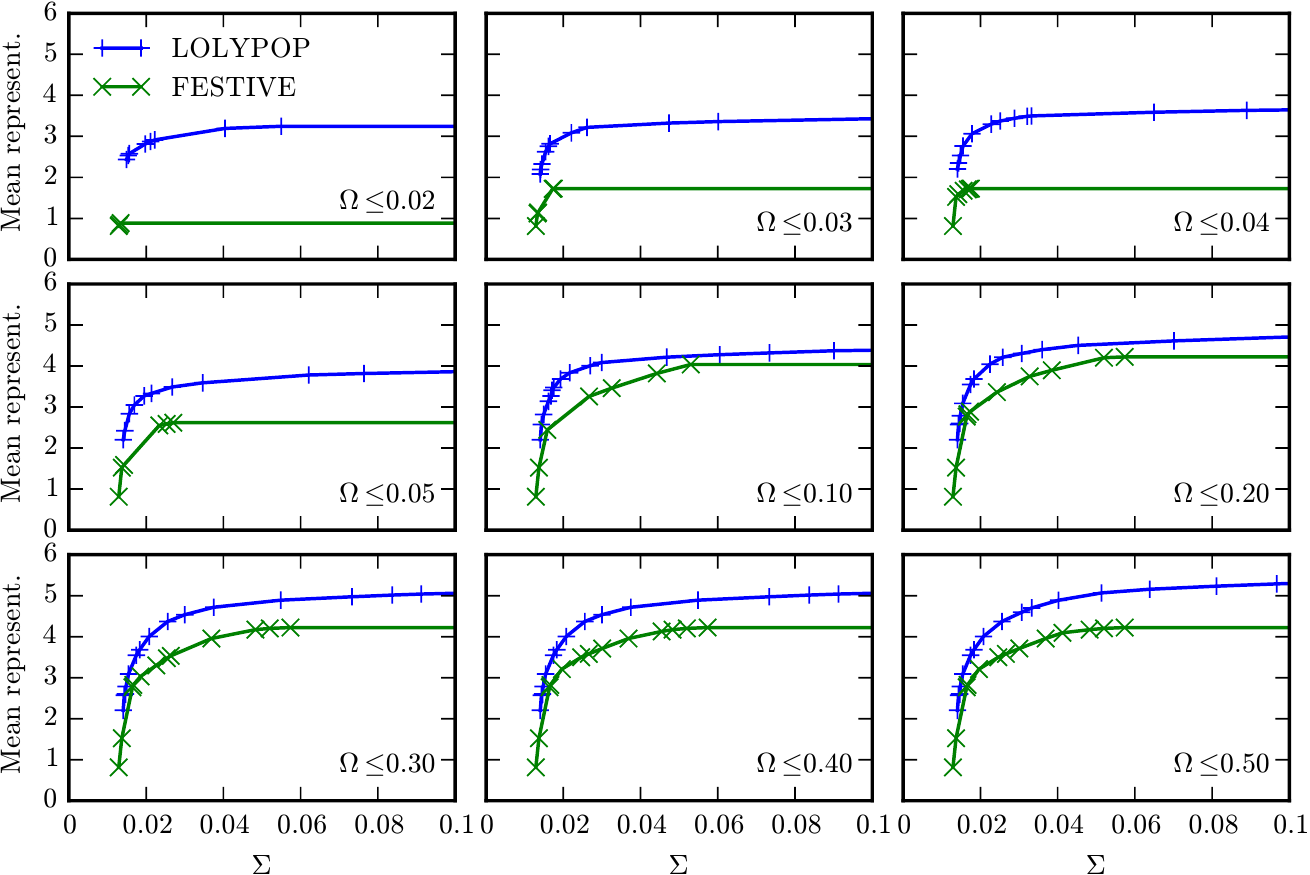}
\caption{Average video quality (representation) as a function of the number of skipped segments $\Sigma$ for different numbers of quality transitions $\Omega$.}
\label{fig:quality_vs_underruns_for_different_switches}
\end{figure}

The fact that $\Sigma$ values below 0.01 were not achieved by either algorithms is in part explained by the throughput outages of several seconds durations contained in several traces. In order to quantify these ``unavoidable'' fractions of skipped segments, we simulate streaming sessions using an adaptation algorithm that always selects the lowest quality. We observed that out of 92 used traces, 66 support streaming at lowest quality without skipped segments. Furthermore, 7 traces have a $\Sigma$ below 0.01, further 10 below 0.05, further 7 below 0.1, one had 0.12 and one has the highest $\Sigma$ of 0.15.

As the main result of our evaluation, we would like to characterize the average video quality as a function of the operating point. Figure~\ref{fig:quality_vs_underruns_for_different_switches} visualizes the average video quality as a function of quality transitions for different numbers of skipped segments. For different values of $\Sigma$ (on the x-axis), we first determine the configuration that (i) achieves the highest average video quality over all traces, (ii) whose average fraction of skipped segments is less or equal to $\Sigma$, and (iii) whose average (relative) number of quality transitions is less then or equal to $\Omega$, where $\Omega\in\left\{\right.$0.02, 0.03, 0.04, 0.05, 0.1, 0.2, 0.3, 0.4, 0.5$\left.\right\}$. The convex hulls of the resulting curves are depicted in Figure~\ref{fig:quality_vs_underruns_for_different_switches}.

We observe that \ac{LOLYPOP} achieves a higher average quality at all operating points. The difference is particularly pronounced for small numbers of quality transitions, where \ac{LOLYPOP} achieves an up to 3 times higher average quality. For high numbers of quality transitions, the difference slightly increases with the number of skipped segments. An interesting observation is that all plots have a more or less pronounced ``knee'', after which the curve goes into saturation and the quality does not increase significantly. In contrast, before the knee, a small increase in the number of skipped segments can bring a huge increase in video quality. In the evaluated network environments, the ``knee'' is typically slightly below $\Sigma=0.02$. In other words, accepting 0.5 to 1 percent more skipped segments, which corresponds to one to two more skipped segments every 400 seconds, can result in an up to twofold improvement in video quality (e.g., for $\Omega\leq 0.2$).

While Figure~\ref{fig:quality_vs_underruns_for_different_switches} shows mean values over all 92 traces, we generated similar plots for each trace individually. In 31 traces, all 9 considered $\Omega$ thresholds resulted in both curves having the same ranges and could thus be compared pointwise. In 21 out of the 31 traces, all 9 curves for LOLYPOP were pointwise strictly greater than the corresponding curves for FESTIVE, while there existed no trace where all 9 curves were pointwise greater for FESTIVE\@. Furthermore, in order to perform a trace-by-trace comparison across all traces, including those in which some curves had different ranges or were intersecting, we compared the integrals of the curves. This comparison revealed that for $\Omega=0.2$, in 53\% of traces, LOLYPOP had a higher integral than FESTIVE; in 38\% of traces, FESTIVE had a higher integral; and in the remaining traces, the values were equal. The corresponding values for $\Omega\in\left\{\right.$0.03, 0.04, 0.05, 0.1, 0.2, 0.3, 0.4, 0.5$\left.\right\}$ are (76\%, 22\%), (82\%, 16\%), (76\%, 22\%), (78\%, 20\%), (86\%, 12\%), (87\%, 11\%), (87\%, 11\%), (89\%, 9\%). We thus observe that for all considered $\Omega$ thresholds, in the majority of traces, the performance of LOLYPOP averaged over the considered range of $\Sigma$ values is higher than the performance of FESTIVE\@.

\begin{figure}
\centering
\includegraphics[scale=1]{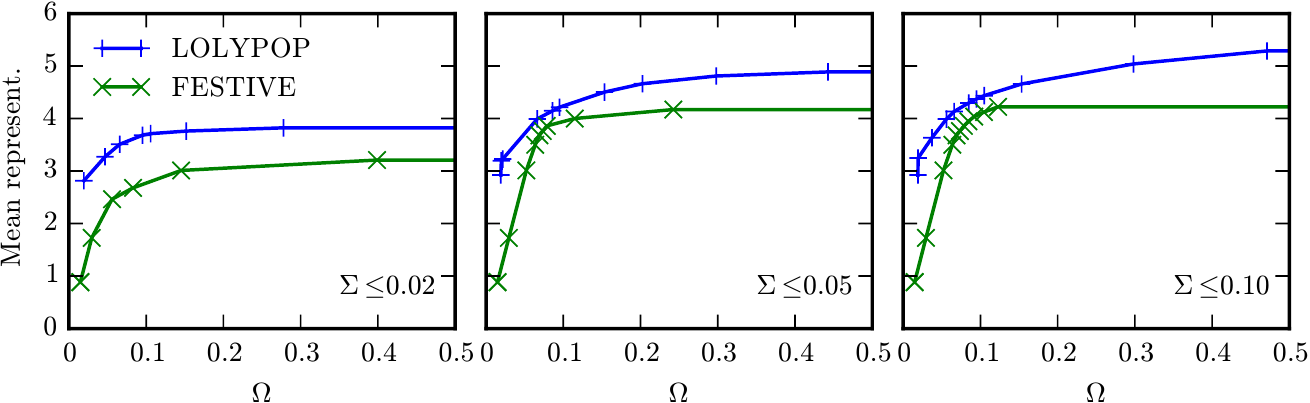}
\caption{Average video quality (representation) as a function of the number of quality transitions $\Omega$ for different numbers of skipped segments $\Sigma$.}
\label{fig:quality_vs_omega_for_different_sigma}
\end{figure}

Similarly to Figure~\ref{fig:quality_vs_underruns_for_different_switches}, Figure~\ref{fig:quality_vs_omega_for_different_sigma} presents plots of quality vs.\ quality transitions for different levels of skipped segments. Here, we observe a similar situtation, albeit the ``knee'' effect is less pronounced in the case of \ac{LOLYPOP} due to the relatively high achieved video quality for low values of $\Omega$.

\begin{figure}
\centering
\includegraphics[scale=1]{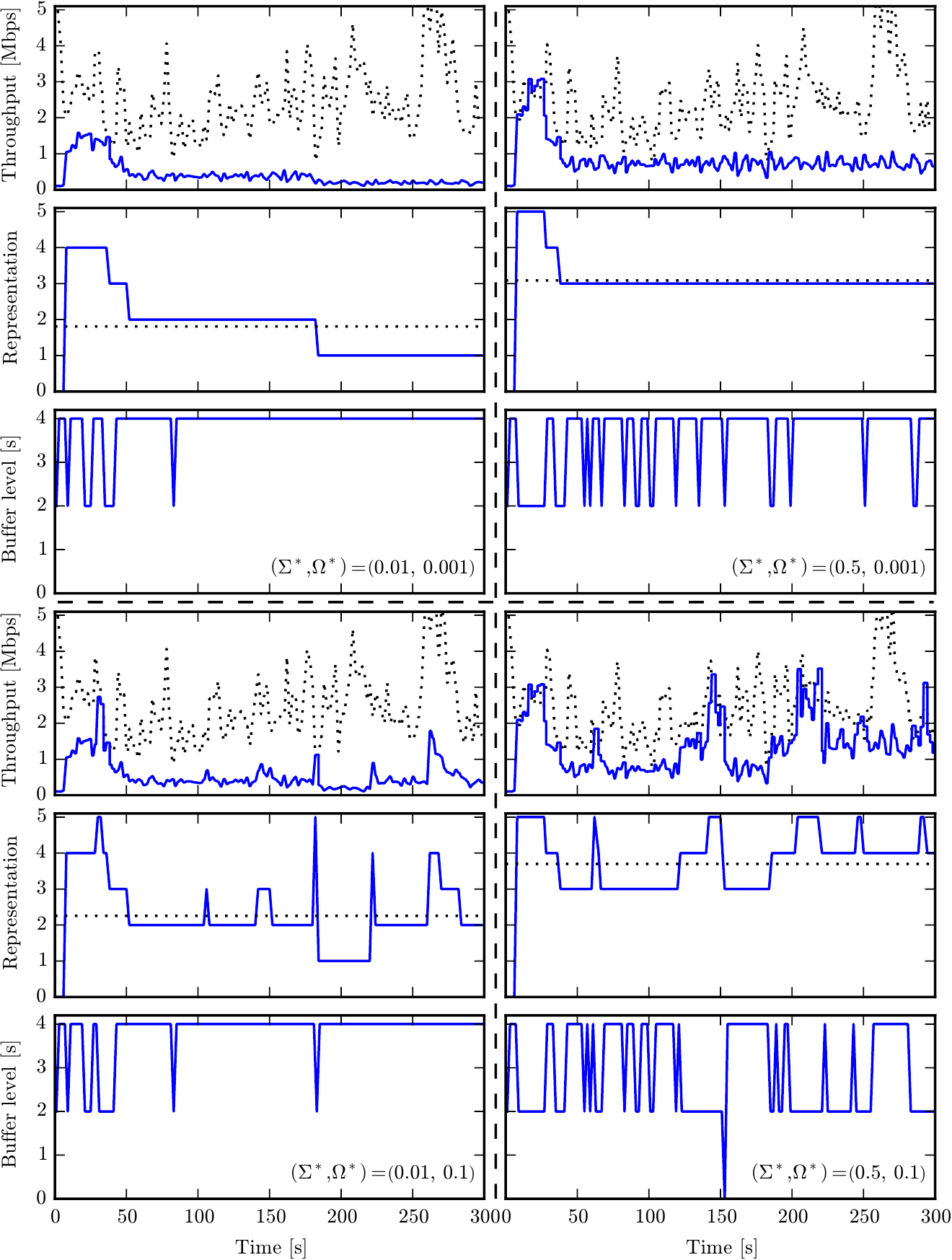}
\caption{Four example runs with different algorithm configurations. See Section~\ref{sec:evaluation_results} for details.}
\label{fig:example_run}
\end{figure}

Finally, Figure~\ref{fig:example_run} depicts four example runs illustrating the behavior of the proposed algorithm with different configurations. For each run, three plots are shown. The top one depicts network throughput and segment \ac{MMBR}. The middle one depicts the representations selected for individual segments and the mean value. The bottom one depicts the buffer level at playback deadlines (a value of 0 results in a skipped segment). In the three upper left subplots, we see a run with low values for both $\Sigma^*$ and $\Omega^*$. Setting $\Omega^*$ to 0.001, we effectively restrict the number of upward transitions to 1, since the whole streaming session has less than 1000 segments. We observe that the algorithm reacts not only to decreased throughput, as seen between seconds 30 and 50, but also to increased uncertainty in throughput dynamics as seen after the strong downward fluctuation at second 170. A higher $\Sigma^*$, as seen in the top right subplots, results in a more aggressive behavior accepting a higher probability for skipping a segment and a higher average quality. The two sets of subplots at the bottom depict runs with $\Omega^*=0.1$. The algorithm is allowed to have more quality transitions, resulting in a further improvement of the average video quality.

\chapter{Conclusion}
\label{sec:conclusion}

In the presented study, we addressed the problem of maximizing \ac{QoE} for \ac{HTTP}-based adaptive low-delay live video streaming. We proposed \ac{LOLYPOP}, an adaptation algorithm designed to operate with a transport latency of a few seconds over wireless access links. \ac{LOLYPOP} leverages predictions of \ac{TCP} throughput distributions on time scales from 1 to 10 seconds.
We studied the performance of several time series prediction methods using IEEE 802.11 traces from various network environments. We observed that the most na\"{\i}ve approach, using the last sample as prediction for the future, has the highest accuracy on all considered time scales. We also observed that the quantiles of the prediction error distribution strongly vary among considered traces requiring a dynamic estimation of the error distribution for each streaming session individually. 
We integrated \ac{LOLYPOP} and a state-of-the-art adaptation algorithm FESTIVE with a streaming client prototype and evaluated them using collected throughput traces. We observed that \ac{LOLYPOP} is able to reach a broad range of operating points, and to outperform the baseline approach w.r.t.\ the mean video quality at each of these operating points by up to a factor of 3.

Our ongoing and future work includes studying the benefits of using cross-layer and context information (i) to adjust the configuration parameters, in particular $\Sigma^*$, in order to achieve a target level of skipped segments, and (ii) to further improve prediction accuracy. Furthermore, we evaluate how $\Delta^p$ can be dynamically tuned in order to achieve minimum latency without dropping the \ac{QoE}. Finally, we plan to evaluate the potential to further reduce latency by reducing the video segment duration.










\acrodef{ADS}{Adobe Dynamic Streaming}
\acrodef{AIC}{Akaike Information Criterion}
\acrodef{AP}{Access Point} \acrodefplural{AP}[AP's]{Access Points}
\acrodef{ARIMA}{Autoregressive Integrated Moving Average}
\acrodef{BOWL}{Berlin Open Wireless Lab}
\acrodef{CDF}{Cumulative Distribution Function} \acrodefplural{CDF}[CDF's]{Cumulative Distribution Functions}
\acrodef{CDN}{Content Delivery Network} \acrodefplural{CDN}[CDN's]{Content Delivery Networks}
\acrodef{CIF}{Context Influence Factor} \acrodefplural{CIF}[CIF's]{Context Influence Factors}
\acrodef{CSS}{Cubic Smoothing Splines}
\acrodef{CV}{Coefficient of Variation}
\acrodef{DASH}{Dynamic Adaptive Streaming over HTTP}
\acrodef{DSL}{Digital Subscriber Line}
\acrodef{DVB}{Digital Video Broadcasting}
\acrodef{ECDF}{Empirical Cumulative Distribution Function} \acrodefplural{ECDF}[ECDF's]{Empirical Cumulative Distribution Functions}
\acrodef{EWMA}{Exponentially Weighted Moving Average}
\acrodef{MPEG-DASH}{Dynamic Adaptive Streaming over HTTP}
\acrodef{GOP}{Group of Pictures}
\acrodef{HAS}{HTTP-Based Adaptive Streaming}
\acrodef{HALS}{HTTP-Based Adaptive Live Streaming}
\acrodef{HD}{High-Definition}
\acrodef{HLS}{Apple HTTP Live Streaming}
\acrodef{HTML}{Hypertext Markup Language}
\acrodef{HTTP}{Hypertext Transfer Protocol}
\acrodef{HIF}{Human Influence Factor} \acrodefplural{HIF}[HIF's]{Human Influence Factors}
\acrodef{IP}{Internet Protocol}
\acrodef{IPTV}{Internet Protocol Television}
\acrodef{ITU}{International Telecommunication Union}
\acrodef{IVP}{Initial Value Problem}
\acrodef{JND}{Just Noticeable Difference}
\acrodef{LAN}{Local Area Network}
\acrodef{LMI}{Linear Matrix Inequality}
\acrodef{LOESS}{Locally Weighted Scatterplot Smoothing}
\acrodef{LOLYPOP}{\textbf{Lo}w-\textbf{L}atenc\textbf{y} \textbf{P}redicti\textbf{o}n-Based Ada\textbf{p}tation}
\acrodef{LOS}{Line-of-Sight}
\acrodef{LQR}{Linear Quadratic Regulator}
\acrodef{LTE}{Long-Term Evolution}
\acrodef{MAC}{Media Access Control}
\acrodef{MANET}{Mobile Ad-Hoc Network}
\acrodef{MCNKP}{Multiple-Choice Nested Knapsack Problem}
\acrodef{MCS}{Modulation and Coding Scheme}
\acrodef{MIMO}{Multiple Input Multiple Output}
\acrodef{MLE}{Maximum Likelihood Estimation}
\acrodef{MMBR}{Mean Media Bit Rate} \acrodefplural{MMBR}[MMBR's]{Mean Media Bit Rates}
\acrodef{MPC}{Model Predictive Control}
\acrodef{MPD}{Media Presentation Description}
\acrodef{MSE}{Mean Squared Error}
\acrodef{MSS}{Microsoft SmoothStreaming}
\acrodef{NAT}{Network Address Translation}
\acrodef{NCS}{Networked Control System}
\acrodef{NLOS}{Non-Line-of-Sight}
\acrodef{NUM}{Network Utility Maximization}
\acrodef{ODE}{Ordinary Differential Equation}
\acrodef{OFDM}{Orthogonal Frequency-Division Multiplexing}
\acrodef{OSI}{Open Systems Interconnection}
\acrodef{OTT}{Over-the-Top}
\acrodef{PI}{Proportional-Integral}
\acrodef{PID}{Proportional-Integral-Derivative}
\acrodef{PSNR}{Peak Signal-to-Noise Ratio}
\acrodef{RTP}{Real-Time Transport Protocol}
\acrodef{RTT}{Round-Trip Time}
\acrodef{TCP}{Transmission Control Protocol}
\acrodef{QoE}{Quality of Experience}
\acrodef{QoS}{Quality of Service}
\acrodef{RMSRE}{Root Mean Square Relative Error}
\acrodef{SES}{Simple Exponential Smoothing}
\acrodef{SIF}{System Influence Factor} \acrodefplural{SIF}[SIF's]{System Influence Factors}
\acrodef{SINR}{Signal-to-Interference-plus-Noise Ratio}
\acrodef{SISO}{Single Input Single Output}
\acrodef{SMA}{Simple Moving Average}
\acrodef{SMC}{Sliding Mode Control}
\acrodef{SNS}{Social Networking Service} \acrodefplural{SNS}[SNS's]{Social Networking Services}
\acrodef{SSI}{Signal Strength Indicator}
\acrodef{SVM}{Support Vector Machine} \acrodefplural{SVM}[SVM's]{Support Vector Machines}
\acrodef{UGC}{User-Generated Content}
\acrodef{URL}{Uniform Resource Locator} \acrodefplural{URL}[URL's]{Uniform Resource Locators}
\acrodef{VBR}{Variable Bit Rate}
\acrodef{VoD}{Video on Demand}
\acrodef{VSC}{Variable Structure Control}
\acrodef{WLAN}{Wireless Local Area Network} \acrodefplural{WLAN}[WLAN's]{Wireless Local Area Networks}
\acrodef{XAP}{Silverlight Application Package}

\bibliography{tomm}

\appendix

\chapter{Information Contained in the Traces}
\label{app:traces}

\begin{figure}[h]
\centering
\includegraphics[scale=1.0]{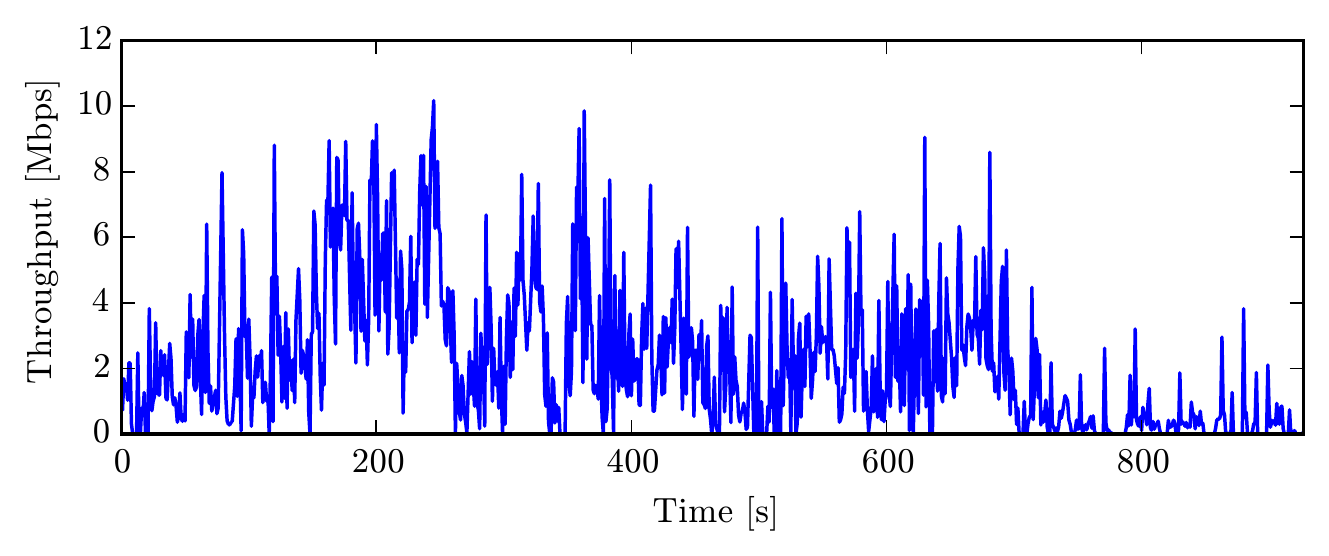}
\label{fig:trace_thrpt_example}
\caption{A complete example trace with a \ac{CV} of 0.879, recorded at a busy outdoor hotspot of a major German telecommunications operator.}
\end{figure}

Each of the collected traces contains several types of information. First, it contains the first 96 bytes of each incoming \ac{TCP} packet belonging to the monitored \ac{TCP} flow. Second, it contains the first 512 bytes of each received IEEE 802.11 frame independent of its destination address, including radiotap headers\footnote{\url{http://www.radiotap.org}} that contain internal \ac{MAC} information, such as the retransmission flag, the \ac{MCS}, and the \ac{SSI}. Except for the radiotap headers, the captured frames are encrypted. Finally, the traces contain periodically recorded values of internal \ac{TCP} variables, obtained using the tcp\_info data structure via the socket interface.
From the traces, we computed time series containing various statistics from overlapping time intervals of 1~s to 10~s duration shifted with a step size of 1~s.
In addition to throughput statistics, we computed statistics of cross-layer information such as for \ac{TCP}: delay jitter statistics and the statistics of outstanding bytes, and, for \ac{MAC}: the number of own frames received, the number of other frames received, \ac{MCS} and \ac{SSI} statistics, and the statistics of retransmissions.
Figure~\ref{fig:trace_thrpt_example} shows the throughput, averaged over one second intervals, of one complete trace with a \ac{CV} of 0.879 recorded at a busy outdoor hotspot of a major German telecommunications operator.

\chapter{Prediction Methods}
\label{app:prediction_methods}



We abbreviate the methods by $\langle$type$\rangle$:$\langle n\rangle$:$\langle$parameters$\rangle$, where $\langle$type$\rangle$ is the name of the method, $n$ is the number of past throughput values used as input, and $\langle$parameters$\rangle$ include further optional configuration parameters.


\section{Simple moving average}

\ac{SMA} is one of the simplest prediction methods. The predicted value is the average over a number of past measurements. The configuration parameters are the number of past measurements and the type of the used mean value: arithmetic, geometric, or harmonic. In the following, we abbreviate this method with SMA:$\langle n\rangle$:$\langle$mean type$\rangle$, where $n\geq 1$ is the number of past measurements, and `mean type' is one of $\{\text{ar},\text{gm},\text{hm}\}$. For example, SMA:2:ar means that the predicted value is the arithmetic mean from two past measurements. In particular, we denote the na\"{\i}ve approach of using the most recent measurement as the predicted value with SMA:1:ar.



\section{Linear extrapolation}

Linear extrapolation is another straightforward prediction method that differs from \ac{SMA} in that it takes into account the linear trend from the past measurements. More specifically, linear extrapolation fits a linear curve into the set of given past measurements, minimizing the \ac{MSE}, and computes the prediction from extrapolating the curve to the prediction horizon. It thus requires at least two past measurements to compute a prediction. We abbreviate linear extrapolation with LinExt:$\langle n\rangle$, where $n\geq 2$ is the number of past measurements.

\section{Double exponential smoothing}

Similar to linear extrapolation, double exponential smoothing tries to account for the trend in the data. In the following, we use a variant of the method, usually referred to as Holt-Winters double exponential smoothing. With Holt-Winters, for the given past measurements $x_1,\ldots,x_n$, the prediction is computed as $x_{n+1}=a_n + b_n$, where $a_n,\;b_n$ are computed by the following recursive procedure.
\begin{align*}
a_n&=\alpha x_n + (1-\alpha)(a_{n-1}+b_{n-1})\,,\;\text{for}\;n>2\\
b_n&=\beta(a_n-a_{n-1})+(1-\beta)b_{n-1}\,,\;\text{for}\;n>2\,,
\end{align*}
with $a_2=x_2$, and $b_2=x_2-x_1$.

The Holt-Winters method has configuration parameters $\alpha$ and $\beta$ that strongly influence the prediction quality and thus have to be carefully tuned. In our work, we tune them for each prediction by minimizing the \ac{MSE} within the past measurements, which is given by $\frac{1}{n-2}\sum_{k=3}^{n}{\left(x_k-(a_{k-1}+b_{k-1})\right)^2}$. Thus, this method requires at least three past values to compute a prediction. As abbreviation, we use HW:$\langle n\rangle$:mse, where $n\geq 3$ is the number of the last values, and \textit{mse} indicates the approach used to tune $\alpha$ and $\beta$.









\chapter{Correlation of Underestimations and Overestimations}
\label{app:correlation_errors}

\begin{figure}[h]
\centering
\includegraphics[scale=1]{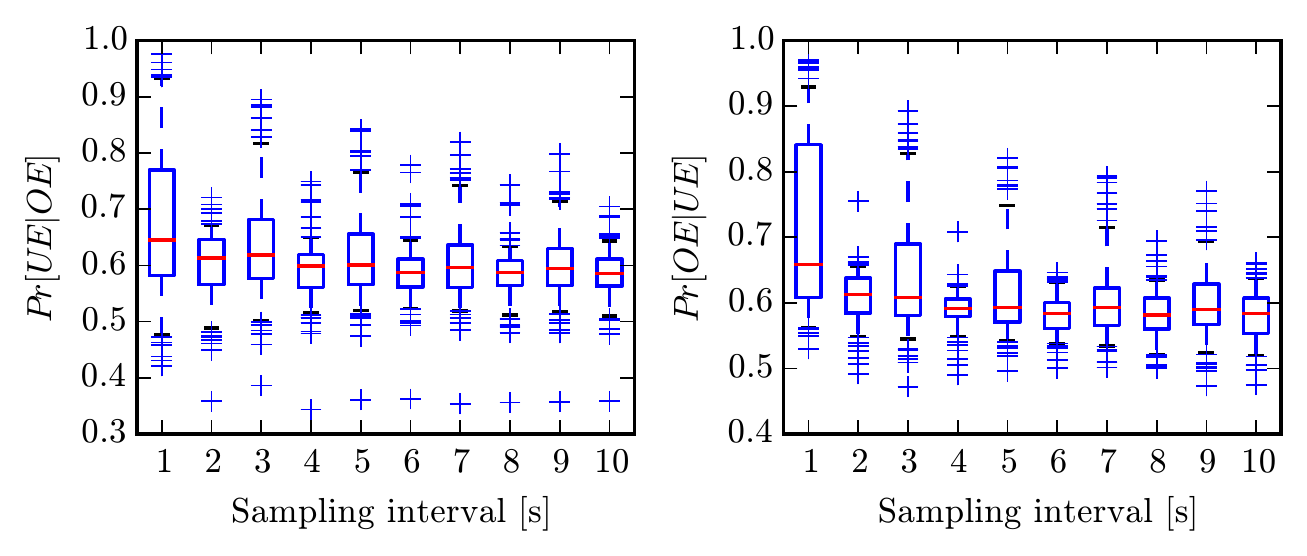}
\caption{Per-trace probability that an underestimation is followed by an overestimation and vice versa. Horizontal line: median, box: quartiles, whiskers: 0.5 and 0.95 quantiles, flier points: outliers. See Section~\ref{sec:performance} for details.}
\label{fig:ue_oe_prob}
\end{figure}

As described in Section~\ref{sec:performance}, we observed that in all the studied traces the probability for occurrences of underestimations and overestimations exhibit significant temporal correlation. In particular, the probability that an underestimation is followed by an overestimation and vice versa is significantly over 50\% for most traces for all time scales, exceeding 80\% or even 90\% in some cases. This observation is directly related to the distinct negative correlation of the throughput process after differencing, as depicted in Figure~\ref{fig:trace_statistics}. The distribution of per-trace values is depicted in Figure~\ref{fig:ue_oe_prob}.

\chapter{Fitting Prediction Error Distributions}
\label{app:fitting_distributions}

For the underestimation errors, distributions are truncated to the range $[0, 1]$, and for the overestimation errors, to the range $[0, \infty)$. The \ac{CDF} $F_{\text{tr}}(\cdot)$ of a distribution truncated to $[a,\,b]$ is obtained from the original \ac{CDF} $F(\cdot)$ as $F_{\text{tr}}(x)=\frac{F(x)-F(a)}{F(b)-F(a)},\;\;x\in[a,\,b]$.

We fit a distribution to the data by minimizing the squared distance ($L^2$-norm) between its \ac{CDF} and the truncated \ac{ECDF}. In order to make the fit more precise in the range which is relevant for adaptive streaming clients, we truncate \acp{ECDF} to the interval $[0.1, 5.0]$ for the overestimation errors, and to the interval $[0.1, 1.0]$ for the underestimation errors. Afterwards, Kolmogorov-Smirnov test is used to verify the goodness of the fit~\cite{Hollander2014}.

\begin{figure}[h]
\centering
\includegraphics{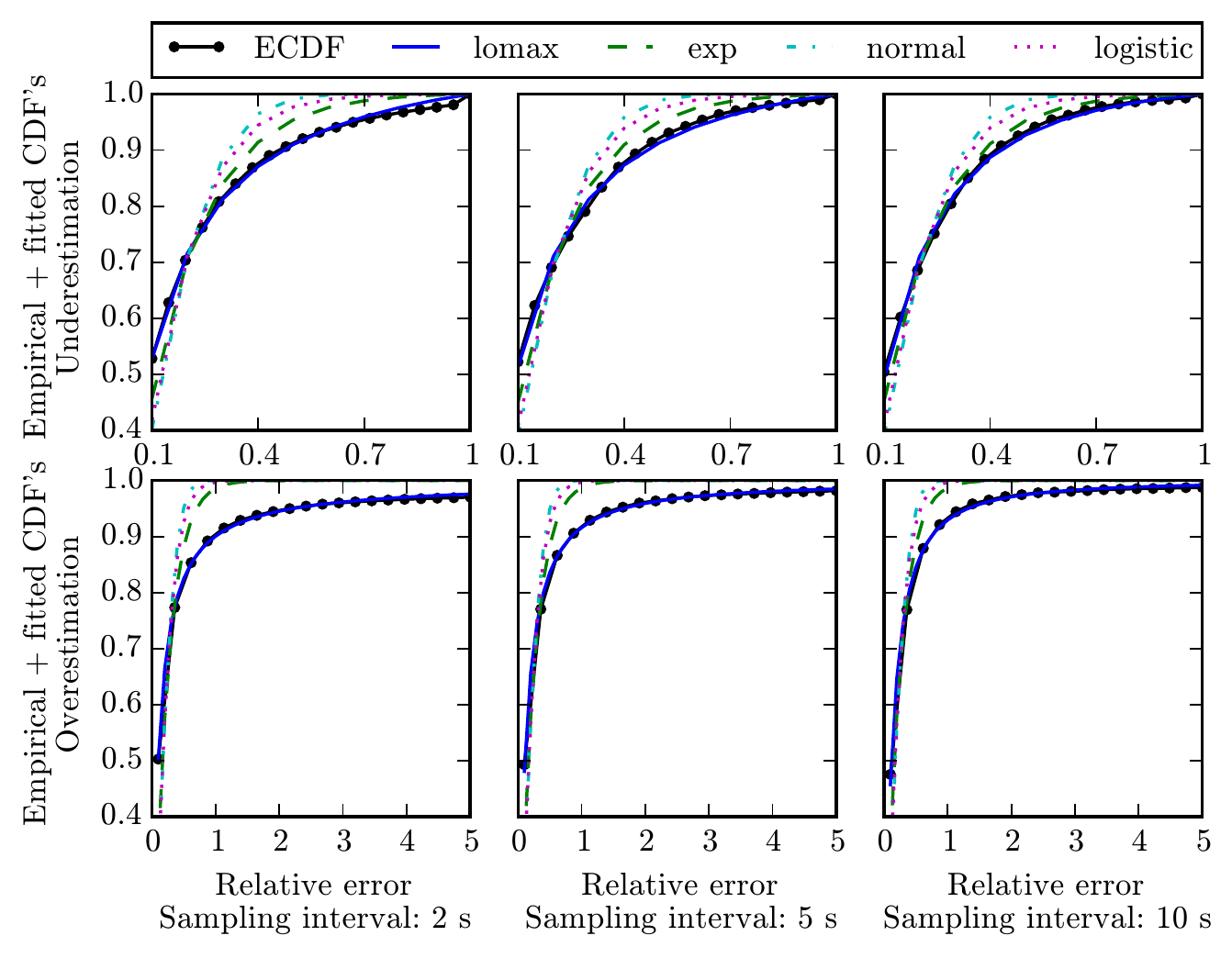}
\caption{Fitting distributions for relative prediction errors. See Section~\ref{sec:modeling} for details.}
\label{fig:distr_fitting_all}
\end{figure}

The results are shown in Figure~\ref{fig:distr_fitting_all}. The \acp{CDF} are fitted to \acp{ECDF} over the joined set of data points from all traces. It turns out that both the underestimation and the overestimation errors are extremely well represented by a Lomax distribution. These findings are consistent with those obtained by fitting the prediction errors from individual traces, which are omitted here.

\end{document}